\pdfoutput=1
\documentclass[11pt,letterpaper]{article}
\usepackage[T1]{fontenc}
\usepackage{lmodern}
\usepackage[margin=1in,letterpaper]{geometry}
\usepackage{amsmath,amssymb,amsthm,mathtools}
\usepackage{microtype}
\usepackage{booktabs}
\usepackage{multirow}
\usepackage{array}
\usepackage[shortlabels]{enumitem}
\usepackage{xcolor}
\usepackage{tikz}
\usetikzlibrary{arrows.meta,positioning,fit,backgrounds,calc,decorations.pathreplacing}
\usepackage{pgfplots}
\pgfplotsset{compat=1.18}
\usepackage{titlesec}
\usepackage{algorithm}
\usepackage[noend]{algpseudocode}

\usepackage[square,authoryear]{natbib}
\let\cite\citep

\usepackage[hidelinks]{hyperref}
\usepackage[nameinlink,capitalize,noabbrev]{cleveref}

\titleformat{\paragraph}[runin]{\normalfont\bfseries}{\theparagraph}{1em}{}

\theoremstyle{plain}
\newtheorem{theorem}{Theorem}[section]
\newtheorem{lemma}[theorem]{Lemma}
\newtheorem{proposition}[theorem]{Proposition}

\newtheorem*{claim}{Claim}
\theoremstyle{definition}
\newtheorem{definition}[theorem]{Definition}
\theoremstyle{remark}

\usepackage{thmtools}

\newcommand{\E}{\mathbb{E}}
\newcommand{\Prb}{\mathbb{P}}
\newcommand{\ind}[1]{\mathbf{1}\!\left[#1\right]}
\newcommand{\Pois}{\mathrm{Pois}}
\newcommand{\Bin}{\mathrm{Bin}}
\newcommand{\OPT}{\mathrm{OPT}}

\newcommand{\dd}{\mathrm{d}}
\newcommand{\NP}{\mathsf{NP}}
\newcommand{\APX}{\mathsf{APX}}
\newcommand{\Ptime}{\mathsf{P}}
\newcommand{\BPP}{\mathsf{BPP}}

\newcommand{\SharpP}{\#\mathsf{P}}
\newcommand{\CountSS}{\textup{\#\textsc{Subset Sum}}}
\newcommand{\PSPACE}{\mathsf{PSPACE}}

\DeclareMathOperator{\Var}{Var}

\definecolor{figblue}{RGB}{40,70,100}
\definecolor{figgreen}{RGB}{38,130,76}
\definecolor{figorange}{RGB}{205,102,12}
\tikzset{
  fig/offline/.style={rectangle,rounded corners=1pt,draw=black!75,
    fill=white,minimum width=8mm,minimum height=6.5mm,inner sep=1pt},
  fig/vtype/.style={circle,draw=figgreen!90!black,fill=figgreen!13,
    text=figgreen!30!black,line width=.6pt,minimum size=7.5mm,inner sep=.5pt},
  fig/etype/.style={circle,draw=figorange!90!black,fill=figorange!15,
    text=figorange!30!black,line width=.6pt,minimum size=7.5mm,inner sep=.5pt},
  fig/note/.style={font=\scriptsize,text=black!60},
  fig/edge/.style={draw=black!55,line width=.6pt},
  fig/vertex/.style={circle,fill=black!75,minimum size=5pt,inner sep=0pt},
  fig/group/.style={rectangle,rounded corners=3pt,draw=black!25,
    fill=black!3,line width=.5pt},
  fig/maps/.style={-{Stealth[length=2.2mm]},draw=black!55,line width=.7pt},
  fig/axis/.style={-{Latex[length=4pt]},draw=black!60,line width=.6pt}
}

\title{Computational Hardness of Online Stochastic Matching}
\author{
  Yifeng Teng\footnote{Email: yifengt@google.com.}\\Google
  \and
  Zongjun Yang\footnote{Email: zy2684@columbia.edu. Part of the work was done
  when the author was a student researcher at Google.}\\Columbia University
}
\date{}

\begin{document}
\maketitle
\thispagestyle{empty}

\begin{abstract}

In the online stochastic matching problem on a bipartite graph, offline vertices are given in advance, and online requests arrive sequentially from known distributions, each matched immediately and irrevocably to a free neighbor or rejected. Computational hardness results of the optimal online strategy have been known only for instances where online vertices have a fixed arrival order \cite{PPSW21}. For random-order arrivals or even i.i.d.\ arrivals, it has been long open since the problem was proposed \cite{FMMM09}.

In this paper, we give the first computational hardness results for online stochastic matching with random-order arrivals. In particular, we prove that for known i.i.d.\ arrivals, computing the online optimum is $\APX$-hard, ruling out a Polynomial-Time Approximation Scheme unless $\Ptime=\NP$, even in unweighted graphs with two offline neighbors per online vertex. For random-order arrivals, even when there is only one offline node (i.e., the prophet secretary setting), computing the optimal online policy is $\SharpP$-hard, resolving an open problem since \cite{DGRTT23}.

\end{abstract}

\clearpage
\setcounter{page}{1}

\section{Introduction}\label{sec:intro}

Online bipartite matching is one of the most fundamental problems in online
algorithms~\cite{KVV90,Meh13,HTW24}. Offline vertices are given in advance,
while online requests arrive one at a time and must be matched immediately and
irrevocably to a free neighbor or rejected. In \emph{online stochastic
matching}, request types are drawn independently from known distributions. The
term originally referred to the \emph{known i.i.d.}\ model~\cite{FMMM09}, in
which all requests follow a single known distribution. More generally, in the
\emph{random-order} model, online requests may come from different known
distributions and arrive uniformly at random; when there is a single offline
vertex, the problem becomes the \emph{prophet secretary}
problem~\cite{EHLM17}. Subsequent work also uses the term when requests with
possibly non-identical known distributions arrive in a \emph{fixed order}
(e.g.,~\cite{PPSW21,BDL22}).

Most of the literature compares polynomial-time algorithms with the
\emph{offline optimum}, which sees all realized requests in advance. Yet a gap
to the offline optimum can come from missing future information rather than
limited computation: even for one offline vertex and two requests in a fixed
order, no online policy obtains more than half of the offline optimum in
expectation~\cite{KS78}, although the optimal online rule is trivial. A growing
line of work instead compares efficient algorithms with the \emph{online
optimum}, the largest expected value of an online policy with unlimited
computation (e.g.,~\cite{PPSW21,BDL22,DGRTT23,NSW23,BDPSW25,STW25}), isolating
the cost of computation. While we know that the online optimum of an online
stochastic matching problem is the value of a Markov decision process that
tracks the free offline vertices and the remaining online requests, the size of
the decision tree is exponential in the input size. Can it be computed, or
approximated within $1\pm\varepsilon$ for every $\varepsilon>0$, in polynomial
time? Such a problem has been open for the random-order case since the original
proposal of the problem~\cite{FMMM09}.

For a fixed arrival order, the answer is negative. \cite{PPSW21} proved that
approximating the online optimum of fixed-order online stochastic matching
within $1\pm c$ for some constant $c>0$ is $\PSPACE$-hard (and hence
$\mathsf{APX}$-hard), and \cite{BDPSW25} showed that this $\PSPACE$-hardness
persists even in unweighted graphs. Related fixed-order hardness results hold
for matroid constraints~\cite{DP24}. All of these reductions come from Stochastic
SAT~\cite{Pap85} and
rely on a fixed arrival order: in the matching reductions, early arrivals
make binary choices, corresponding to variable assignments, and later arrivals
test them, corresponding to clauses of a SAT formula, so every test arrives
after the choices it inspects. Under random-order arrivals, an instance cannot
schedule choices before tests, and under known i.i.d.\ arrivals, request types
also repeat. This means that previous reductions cannot be directly applied to
random-order instances.
Even when there is a single offline vertex, while \cite{DGRTT23} proved that the prophet secretary problem admits a PTAS algorithm, it remains open whether computing the optimal strategy is $\NP$-hard.
In fact, even beyond matching, we are not aware of \emph{any}
computational hardness result for \emph{any} online resource allocation problem
with a tractable offline counterpart under random-order arrivals, let alone for
i.i.d.\ arrivals.
We defer more related work to \cref{sec:related}.

\subsection{Our results}\label{sec:contribution}

In this paper, we show that a random arrival order does not make the online
optimum tractable, establishing the first computational hardness results under
random-order arrivals. Our main result is that
approximating the online optimum of online bipartite matching is
$\mathsf{APX}$-hard in the known i.i.d.\ model~\cite{FMMM09}, and hence also
under general random-order arrivals.

\begin{theorem}[Informal version of \cref{thm:main}]\label{thm:intro-matching}
Approximating the online optimum of known i.i.d.\ matching is
$\mathsf{APX}$-hard, even for unweighted
instances where each request type has two offline neighbors.
\end{theorem}

The APX-hardness result rules out the existence of any PTAS (unless $\Ptime=\NP$) for online stochastic matching with i.i.d.\ arrival.  Our second result is that in the
random-order case, even when there is a single offline node, computing an optimal policy is
$\SharpP$-hard even though a PTAS is known~\cite{DGRTT23}.

\begin{theorem}[Informal version of \cref{thm:decisions}]\label{thm:intro-secretary}
In prophet secretary, computing an optimal online policy is $\SharpP$-hard.
\end{theorem}

The result holds even when every variable has a binary (0 or 1) value except for two deterministic values in $(0,1)$. Because single-item instances under fixed-order or known i.i.d.\ arrivals can be
solved in polynomial time by backward induction, this hardness comes from the
random arrival order alone.

\subsection{Technical overview}\label{sec:techniques}

\paragraph{Known i.i.d.\ matching.}
We reduce from approximating the maximum cut size $C(G)$ of a cubic graph
$G$~\cite{AK00} (we only need a connected graph with $O(1)$ degree). From an
$n$-vertex input graph $G$, we build an unweighted online bipartite matching
instance whose offline side has $2n$ vertices with a pair
$u^+,u^-$ for each vertex $u$ of $G$. The online side has slightly more
than $n$ i.i.d.\ arrivals with two classes of types. Almost all arrivals have
vertex types, where each vertex $u$ of $G$ gives a type $u$ that can be matched to
$\{u^+,u^-\}$, and arrives roughly once in expectation. Each edge $e=(u,v)$ of
$G$ also gives two rare edge types $e^+$ and $e^-$, with neighbors
$\{u^+,v^+\}$ and $\{u^-,v^-\}$, each of which arrives far less than once in
expectation.
As the instance is unweighted, there exists a greedy optimal policy, so the choices reduce to breaking ties between the two
neighbors.
At the first vertex arrival of $u$, a greedy policy must match
either $u^+$ or $u^-$, and the sign of the offline vertex left unmatched places $u$
on side $+$ or $-$ of a cut of $G$. Once $u$ and $v$ have each had one vertex
arrival, one offline vertex in $\{u^+,u^-\}$ and one in $\{v^+,v^-\}$ remain
free. If they have different signs (say $u^+$ and $v^-$), either of $e^+$ and
$e^-$ can be matched, whereas if they have the same sign (say $u^+$ and
$v^+$), only one of $e^+$ and $e^-$ has a free neighbor. Cutting $e$ in $G$
thus doubles the chance of matching an edge arrival of $e$.

In a fixed
arrival order, this construction already gives $\APX$-hardness. Suppose
that each vertex type arrives deterministically once first, and then for each
of the $m$ edges $e$, an edge arrival comes independently with a small
constant probability $\lambda$, equally likely to have type $e^+$ or $e^-$.
Since the first $n$ arrivals are deterministic, a greedy policy
gets $n$ matches from them and fixes a single cut of $G$. The edge $e$ then
gives a match with probability $\lambda$ if it is cut and $\lambda/2$ if it is
uncut, unless an earlier edge arrival sharing an endpoint with $e$ has already
matched the free offline vertex there. Since $G$ is cubic, two edge arrivals
sharing a vertex of $G$ both arrive with probability only $O(\lambda^2)$ per
edge, losing at most $O(\lambda^2 m)$ expected matches over all $m$ edges.
Maximizing the expected matching size therefore amounts to choosing a maximum
cut of $G$, and the online optimum equals $n+\lambda m/2+\lambda C(G)/2$ up to
an $O(\lambda^2 m)$ error. We take $\lambda$ to be a small constant, so that
the $\APX$-hardness of approximating $C(G)$ implies the $\APX$-hardness of
approximating the online optimum.

Turning this construction into an i.i.d.\ reduction faces two challenges that
do not exist in a fixed arrival order. First, edge arrivals come at arbitrary
times. When an edge arrival of $(u,v)$ comes, $\{u^+,u^-\}$ and $\{v^+,v^-\}$
may each have $0$ to $2$ free offline vertices, and matching $u^+$ or $u^-$
early adds nothing to the final matching size if later vertex arrivals of $u$
would have filled $\{u^+,u^-\}$ anyway. In the analysis, we replace the fixed
number of i.i.d.\ arrivals by a Poisson number of arrivals with the same mean,
as is common for i.i.d.\ arrivals (e.g.,~\cite{HS21}). This changes the online
optimum by only an $o(1)$ fraction, and the arrivals of the different types
become independent Poisson processes over $[0,1]$. We also block every edge
arrival that shares an endpoint with an earlier edge arrival, which costs
$O(\lambda^2m)$ as in a fixed order (\cref{cor:transfer}). In
this adjusted Poisson model, the probability that matching $u^+$ or $u^-$
increases the final matching size depends only on the arrival time and on how
many of $\{u^+,u^-\}$ are free (\cref{lem:accounting}). Since a pair with $0$
or $2$ free offline vertices is symmetric between $+$ and $-$, an edge $(u,v)$
rewards different signs only when $u$ and $v$ have each had one vertex arrival
so far, while all other cases give the same expected gain regardless of the
cut (\cref{lem:cutvalue}).

Second, unlike the
deterministic vertex arrivals in a fixed arrival order, i.i.d.\ arrivals let a
policy choose the side of $u$ adaptively at the first vertex arrival of $u$,
based on earlier arrivals. At any time, it can therefore cut the random
subgraph induced by the vertices that have had exactly one vertex arrival so
far, which on $G$ can strictly beat any fixed cut (for instance, on a
triangle, any two vertices induce a single edge and are bipartite). To
eliminate this adaptivity gain, we apply the reduction not to $G$ itself,
but to its blow-up $\widetilde G$ (\cref{def:blowup}), which
replaces each vertex by $B$ copies and each edge by a $B\times B$ complete
bipartite graph. By concentration analysis, we show that every vertex of $G$ has nearly the same
fraction of its $B$ copies in that random subgraph, so even an adaptive policy
beats a fixed maximum cut by at most an $O(1/\sqrt{B})$ fraction of the cut
term (\cref{lem:blowup,lem:gap}). We take $B$ to be a sufficiently large
constant and $\lambda$ to be a small constant multiple of $1/B$. Then the
online optimum becomes an explicit affine function of $C(G)$ whose cut term is
a constant fraction of the online optimum and dominates all error terms, which
proves $\APX$-hardness for known i.i.d.\ matching.

\paragraph{Prophet secretary.}
We reduce from \CountSS~\cite{FH09}, which asks for the number of solutions
$S\subseteq[n]$ with $\sum_{i\in S}a_i=b$, given positive integers
$a_1,\dots,a_n$ and a target $b$. After scaling $a_1,\dots,a_n$ and $b$ down by
an exponentially large
common factor so that $\sum_{i=1}^n a_i$ is
small, we consider an instance with an online vertex of
deterministic value $x\in(0,1)$ and $n$ online vertices with values in
$\{0,1\}$, which have value $1$ with probabilities $a_1,\dots,a_n$. An optimal
policy always accepts $1$ and rejects $0$; when $x$ arrives with no earlier
$1$, it accepts $x$ if $r_S<x$ and rejects $x$ if $r_S>x$, where
$r_S=1-\prod_{i\in S}(1-a_i)\approx\sum_{i\in S}a_i$ is the probability that the
set $S$ of remaining online vertices contains a value $1$. Increasing $x$
slightly raises the expected reward only on the realizations where the policy
accepts $x$, so the slope of the online optimum with respect to $x$ is the
probability of accepting $x$, namely the probability that no earlier $1$ arrived
and $r_S<x$ (\cref{fig:slope}). Because each $r_S$ is extremely close to
$\sum_{i\in S}a_i$, the jump in this slope across a narrow interval around $x=b$
equals the probability that no earlier $1$ arrived and the random remaining
set $S$ satisfies $\sum_{i\in S}a_i=b$.

Under random-order arrivals, every $k$-element subset has the same probability
$1/((n+1)\binom nk)$ of being the remaining set after $x$, and with our
exponential scaling an earlier $1$ arrives with exponentially small probability.
Once we preprocess \CountSS{} so that all solutions have the same size $k$,
that slope jump around $b$ is proportional to the number of solutions up to
negligible error, so evaluating the online optimum at three points around $b$
and rounding recovers the count exactly (\cref{lem:counting}). This makes
computing the online optimum of the instance with a single non-binary value $\SharpP$-hard (\cref{lem:opt}). In this
instance, an optimal policy is still easy to compute, because by the time $x$
arrives the remaining set $S$ is already known and $r_S$ can be computed
directly. Adding a second online vertex of deterministic value $q\in(0,1)$
makes computing an optimal policy $\SharpP$-hard as well
(\cref{thm:decisions}).

\section{Model}\label{sec:model}

We describe the online stochastic matching problem formally, including the random-order arrival and i.i.d.\ arrival settings.

An instance $I$ of online stochastic matching has a finite set $U$ of offline
vertices, a finite set of types, and $T$ online vertices. The type $y$ of an online vertex specifies
the weight $w(y,u)\geq0$ for each offline vertex $u$, and the neighbors of $y$
are the offline vertices $u$ with $w(y,u)>0$. Each online vertex $j$ has a
random type, drawn from a known distribution $\mathcal D_j$ independently of
the other online vertices.

We consider \textit{random-order arrivals}, where the online vertices arrive one at a
time in a uniformly random order, independent of their types. When an online
vertex arrives, it reveals its type $y$ and its identity, and an online policy either matches it
to a free neighbor $u$ of $y$ and gains reward $w(y,u)$, or
rejects it. Each offline vertex can only be matched once, and the decisions are irrevocable.
An online policy can make decisions based on the online vertices that have arrived so far and their matching history, but it does not see future arrivals.

We measure a policy by its expected total weight of the matched edges; let $M_\pi$ be the total weight that a policy $\pi$ obtains.

\begin{definition}[Online optimum]\label{def:opt}
The online optimum of an instance $I$ is
$\OPT(I)=\sup_\pi\E[M_\pi]$,
where the supremum is over all policies $\pi$ for $I$. An online policy $\pi$ is
optimal if $\E[M_\pi]=\OPT(I)$.
\end{definition}

The online optimum is the optimal value of a Markov decision process with
exponentially many states. We study the complexity of two tasks on a class of
instances.
\begin{itemize}[leftmargin=*,itemsep=1pt]
\item \textit{Approximating the online optimum within a factor
$1\pm\varepsilon$}: Given an instance $I$ of the problem, find a
number $\bar V$ with $|\bar V-\OPT(I)|\leq\varepsilon\OPT(I)$.
\item \textit{Computing an ($\varepsilon$-approximately) optimal policy}: Given an instance $I$ of the problem, find a policy that achieves the online optimum (within a factor of $1-\varepsilon$).
\end{itemize}

\subsection{Unweighted i.i.d.\ matching}\label{sec:iid}

We consider the unweighted, known i.i.d.\
model~\cite{FMMM09}, where all weights of the online types are $0$ or $1$ and all online vertices
are generated independently from the same distribution $\mathcal D$ of types. Then $M_\pi$ is the number
of matches of $\pi$.
Since all distributions are the same, the \textit{i.i.d.} model is a simple special case of random-order arrivals.
For $\ell\geq0$, we let $y_1, \cdots, y_\ell$ be the types of the first $\ell$ online vertices, and let $M_\pi(\ell)$ be the number of
matches obtained from these vertices. We will refer to $\E[M_\pi(T)]$ as the value of $\pi$. Since there are finitely many types,
there exists a deterministic policy that is optimal,
so $\OPT(I)=\max_\pi\E[M_\pi(T)]$ where the maximum is over deterministic policies $\pi$.

In the unweighted setting, it suffices to consider only greedy policies, i.e., policies that match
an online arrival whenever there is a free neighbor.

\begin{proposition}[Greedy policies]\label{prop:greedy}
For every deterministic policy $\pi$, there is a deterministic greedy policy
$\pi'$ such that, on every sequence of types, $M_{\pi'}(\ell)\geq M_\pi(\ell)$
for every $\ell\geq0$.
\end{proposition}

\subsection{Prophet secretary}\label{sec:ps}

We will also consider prophet secretary~\cite{EHLM17}, the special case with a
single offline vertex $u$. Let $X_j=w(y,u)$ be the value of an online
vertex $j$ with type $y$, and the question is to maximize the selected value in expectation.
In the random-order model, $X_j$ are independent random variables
with known distributions, but arrive in a uniformly random order.

In the prophet secretary setting, an optimal policy is given by backward induction over the set of online
vertices still to come.

\section{$\APX$-Hardness of i.i.d.\ Matching}\label{sec:matching}

In this section, we prove the $\APX$-hardness of approximating the online
optimum for i.i.d.\ matching.

\begin{restatable}{theorem}{mainthm}\label{thm:main}
For online stochastic matching with i.i.d.\ arrivals, there is a constant
$\varepsilon_0>0$ such that approximating the online optimum within a factor
$1\pm\varepsilon_0$ is $\NP$-hard, even when the instance is unweighted and
every type has exactly two neighbors.
\end{restatable}

We reduce from the problem of approximating the size of the maximum cut of
simple cubic graphs, i.e., graphs in which all vertices have degree exactly
$3$. For a graph $G=(V,E)$, a cut of the graph is a mapping
$\sigma\colon V\to\{+,-\}$, and the size of the cut is $|\delta_G(\sigma)|$,
where $\delta_G(\sigma)=\{(u,v)\in E:\sigma(u)\neq\sigma(v)\}$ is the set of
edges that $\sigma$ cuts. We let $C(G)$ be the
maximum size of a cut in graph $G$.

\begin{proposition}[\cite{AK00}]\label{thm:maxcut}
There is a constant $\eta_0\in(0,1)$ such that approximating $C(G)$ within a
factor $1\pm\eta_0$ is $\NP$-hard, even when $G$ is a simple cubic graph.
\end{proposition}

We will construct an online stochastic matching instance from $G$, such that
any approximation of the online optimum will lead to an approximation of
$C(G)$.

\subsection{From cut to online matching}\label{sec:construction}

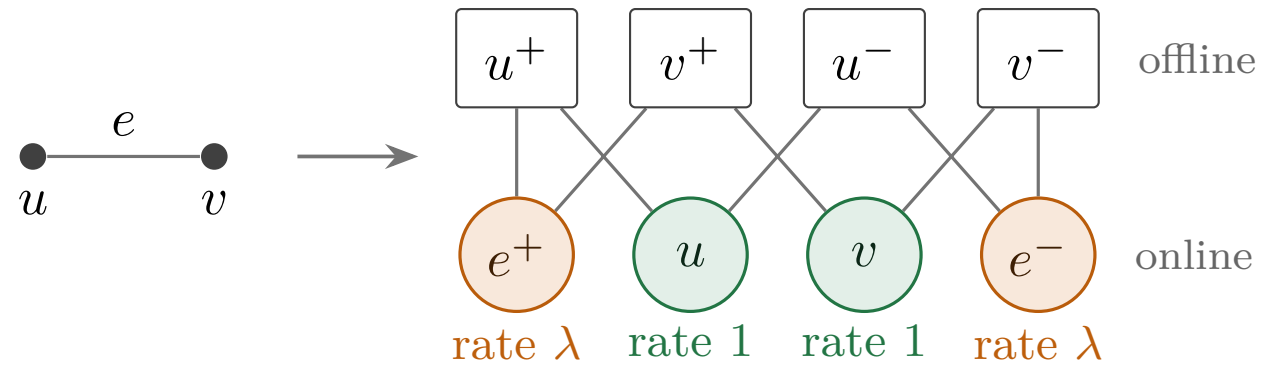
\begin{figure}[t]
\centering
\begin{tikzpicture}[font=\small,
    rate/.style={font=\footnotesize,anchor=north,inner sep=1pt}]
  \node[fig/vertex,label=below:$u$] (gu) at (-3.2,0) {};
  \node[fig/vertex,label=below:$v$] (gv) at (-2,0) {};
  \draw[fig/edge] (gu)--node[above] {$e$}(gv);
  \draw[fig/maps] (-1.45,0)--(-.65,0);
  \node[fig/offline] (up) at (0,.65) {$u^+$};
  \node[fig/offline] (vp) at (1.15,.65) {$v^+$};
  \node[fig/offline] (um) at (2.3,.65) {$u^-$};
  \node[fig/offline] (vm) at (3.45,.65) {$v^-$};
  \node[fig/etype] (ep) at (0,-.65) {$e^+$};
  \node[fig/vtype] (u) at (1.15,-.65) {$u$};
  \node[fig/vtype] (v) at (2.3,-.65) {$v$};
  \node[fig/etype] (em) at (3.45,-.65) {$e^-$};
  \node[fig/note,anchor=west] at ($(vm.east)+(.12,0)$) {offline};
  \node[fig/note,anchor=west] at ($(em.east)+(.12,0)$) {online};
  \foreach \t in {ep,em}
    \node[rate,text=figorange!92!black] at ($(\t.south)+(0,-.04)$) {rate $\lambda$};
  \foreach \t in {u,v}
    \node[rate,text=figgreen!90!black] at ($(\t.south)+(0,-.04)$) {rate $1$};
  \begin{scope}[on background layer]
    \draw[fig/edge] (ep)--(up) (ep)--(vp) (u)--(up) (u)--(um)
      (v)--(vp) (v)--(vm) (em)--(um) (em)--(vm);
  \end{scope}
\end{tikzpicture}
\caption{The part of $I(G,\lambda)$ built from an edge $e=(u,v)$ of $G$.}
\label{fig:instance}
\end{figure}

For any simple graph $G=(V,E)$ with $|V|=n$, $|E|=m$ and largest degree $\Delta_G$, and any parameter
$\lambda>0$, we construct an online stochastic matching instance
$I(G,\lambda)$ with $2n$ offline vertices and $n+2m$ possible arrival types
(\cref{fig:instance}).
\begin{itemize}[leftmargin=*,itemsep=1pt]
\item \textit{Offline vertices}: For each $u\in V$, we construct a pair of
offline vertices $u^+,u^-$.
\item \textit{Vertex type arrivals}: For each $u\in V$, we construct an online
arrival type $u$ with neighbors $\{u^+,u^-\}$, with rate $\lambda_u=1$.
\item \textit{Edge type arrivals}: For each $e=(u,v)\in E$, we construct two
online arrival types $e^+$ and $e^-$, with the neighbors being $\{u^+,v^+\}$
and $\{u^-,v^-\}$ respectively. Each type has rate
$\lambda_{e^+}=\lambda_{e^-}=\lambda$.
\end{itemize}
There are $T=\lceil R\rceil$ online vertices, where $R=n+2m\lambda$ is the total
arrival rate. The probability of each type is proportional to its rate, that
is, $1/(n+2m\lambda)$ for each vertex type, and $\lambda/(n+2m\lambda)$ for
each edge type. We will choose the parameter $\lambda$ to be small, so that the
vertex types make up most of the matching and create a baseline value, while the policy receives bonuses for matching the rare edge type arrivals.

We consider the deterministic online matching policy $\pi_\sigma$ induced by a
given cut $\sigma$ of $G$. The policy will always match an online arrival when
it has a free neighbor, and uses the cut $\sigma$ to break ties when
matching vertex type arrivals. Specifically, for a vertex arrival of type $u$
with both neighbors $u^+,u^-$ free, the policy matches it to the offline
vertex $u^{-\sigma(u)}$. When matching an edge arrival, it prefers an offline
vertex $w^s$ such that $w^{-s}$ is also free. See \cref{alg:cut-policy}
for the details of the policy $\pi_\sigma$.

\begin{algorithm}[ht]
\caption{The policy $\pi_\sigma$ induced by a cut $\sigma$ of $G$.}
\label{alg:cut-policy}
\begin{algorithmic}[1]
\For{$k=1,2,\dots$}
  \If{$y_k=u$ for some $u\in V$}
    \If{$u^+$ and $u^-$ are both free}
      \State Match the $k$-th online arrival to $u^{-\sigma(u)}$.
    \ElsIf{exactly one of $u^+$ and $u^-$ is free}
      \State Match the $k$-th online arrival to this vertex.
    \Else
      \State Reject the $k$-th online arrival.
    \EndIf
  \EndIf
  \If{$y_k=e^s$ for some $e=(u,v)\in E$ and $s\in\{+,-\}$}
    \If{$w^s$ and $w^{-s}$ are both free for some $w\in\{u,v\}$}
      \State Match the $k$-th online arrival to $w^s$ for the first such $w$ in
      a fixed order of $V$.
    \ElsIf{$w^s$ is free for some $w\in\{u,v\}$}
      \State Match the $k$-th online arrival to $w^s$ for the first such $w$ in
      a fixed order of $V$.
    \Else
      \State Reject the $k$-th online arrival.
    \EndIf
  \EndIf
\EndFor
\end{algorithmic}
\end{algorithm}

An important relationship we will show later is that a larger cut $\sigma$
will result in more expected matches under \cref{alg:cut-policy}. We will then show that the policy induced by a maximum
cut is approximately optimal, once the graph is blown up, and give an
invertible mapping from the size of the maximum cut to the approximate value of
the optimal online matching policy.

\subsection{Estimating via Poisson arrivals}\label{sec:estimate}
We estimate the matching value of the policy $\pi_\sigma$ by its value under a
different but similar Poisson arrival model. The Poisson model makes the
matching value easier to characterize, and we bound the difference of values of
any online policy under the two models (\cref{lem:transfer}).

\paragraph{Poisson arrival model with mild adjustments.}
We consider a Poisson process of rate $R$; let $\tau_1<\tau_2<\cdots$ be its
points, and let $K(t)$ be the number of points in $[0,t]$. The type of each
arrival is independently generated from the same distribution as in
$I(G,\lambda)$. The online arrivals are all points in $[0,1]$, so the number of
online arrivals is $K(1)$.
The arrival of different online types can be considered as independent Poisson
processes with their respective rates.
The Poisson process is used to generate the $K(1)$ arrivals; the online policy sees the type of each arrival, but not their underlying arrival time in the Poisson process.
We note that Poisson arrivals have appeared in previous work on
online stochastic matching~\cite{HS21}.

For the edge types in the Poisson arrival, we make a mild adjustment to avoid
interference in later analysis; the adjustment happens with a rare event, so the readers can also consider this as dropping a small-probability event. If the edge of an edge arrival shares an
endpoint with the edge of any previous edge arrival, we block it from the
Poisson arrival, or equivalently, force any policy to reject it by setting its
set of neighbors to be empty. Specifically, after an online arrival with type
$e^s$ ($s\in\{+,-\}$, $e=(u,v)\in E$), we will block any future edge arrival with
type $f^+$ or $f^-$ when $f$ is an edge that is incident to $u$ or
$v$; the rule applies even when $e^s$ itself is blocked. The formal setup of
the adjusted Poisson arrival model can be found in \cref{alg:adjusted}.

\begin{algorithm}[ht]
\caption{The Poisson arrival model for $I(G,\lambda)$, with mild adjustment.}
\label{alg:adjusted}
\begin{algorithmic}[1]
\State Generate points $\tau_1<\tau_2<\cdots$ of a Poisson process of rate
$R$.
\State Let $K(1)=|\{\ell:\tau_\ell\in[0,1]\}|$ be the number of online
arrivals.
\State $S\gets\varnothing$.\Comment{endpoints of the earlier edge arrivals}
\For{$k=1,\dots,K(1)$}
  \State Independently generate the type $y_k$ to be $y$ with probability
  $\lambda_y/R$.
  \If{$y_k=e^+$ or $e^-$ for some $e=(u,v)\in E$}
    \If{$u\in S$ or $v\in S$}
      \State Set the neighbors of the $k$-th online arrival to be $\varnothing$.
    \EndIf
    \State $S\gets S\cup\{u,v\}$.
  \EndIf
\EndFor
\end{algorithmic}
\end{algorithm}

We next bound the value of any fixed policy under the standard i.i.d.\ and the  Poisson arrival models. Given a deterministic online matching policy
$\pi$, let $\widehat M_\pi$ be its number of matches in the adjusted Poisson
arrival model, and refer to its expectation as the \textit{Poisson value} of the policy.

\begin{restatable}[The adjusted Poisson model]{lemma}{transferlemma}\label{lem:transfer}
For every deterministic policy $\pi$ for $I(G,\lambda)$,
\[
 \E[M_\pi(T)]-4\Delta_G\lambda^2m-\sqrt R-1\ \leq\ \E[\widehat M_\pi]\ \leq\
 \E[M_\pi(T)]+\sqrt R+1.
\]
\end{restatable}

To prove \Cref{lem:transfer}, an important observation is that blocked
edges are rare, with a quadratic rate in $\lambda$. Indeed, an edge arrival type could only block $4\Delta_G$ potential future types, and each edge arrival type has a small arrival rate of $\lambda$.

\Cref{lem:transfer} further gives us the bound on the differences between optimal values under the two arrival models. Let $\widehat\OPT=\sup_\pi\E[\widehat M_\pi]$ be the optimal Poisson value over all online policies.

\begin{restatable}{corollary}{transfercor}\label{cor:transfer}
The online optimum of $I(G,\lambda)$ satisfies
\[
 \widehat\OPT-\sqrt R-1\ \leq\ \OPT(I(G,\lambda))\ \leq\
 \widehat\OPT+4\Delta_G\lambda^2m+\sqrt R+1.
\]
\end{restatable}

\paragraph{Net gains of matching edge type arrivals.}
Since most arrivals are vertex type arrivals, we consider them as the baseline
for the Poisson value. If no edge type arrival is matched, the Poisson value of
a greedy policy is $n\cdot\E[\min\{\Pois(1),2\}] = (2-3/e) n$, since there are two offline
vertices $u^+,u^-$ for each $u\in V$, and each vertex type has rate
$\lambda_u=1$.

We then consider: \textit{on top of this baseline, what is the net gain of
matching an edge type arrival?} For example, let $d$ be the number of free
vertices in $\{u^+,u^-\}$, and at time $t$ an edge arrival $e^s$ is matched to $u^s$.
In the case where there are at least $d$ later vertex
arrivals of type $u$, matching $e^s$ does not increase the total number of
successful matchings since it will displace a later vertex arrival; otherwise
matching $e^s$ will create a net gain of $1$ in the eventual matching. We
formally define this \textit{net gain} in expectation:
\begin{equation*}
  h_d(t)=\Prb(\Pois(1-t)\leq d-1)=\begin{cases}
    0 & (d=0),\\
    e^{t-1} & (d=1),\\
    (2-t)e^{t-1} & (d=2).
  \end{cases}
\end{equation*}

We can then characterize the total net gains that a policy collects in expectation.
Conditional on a time-$t$ arrival of edge type $e^s$, with $e=(u,v)$, we
consider the events: (I) the current arrival will not be blocked by the arrival model,
and (II) the policy matches the
arrival to some $w^s$, $w\in\{u,v\}$, where $d$ of $\{w^+,w^-\}$ are free.
We let $P_e(t)$ be the probability of (I), and $Q^s_{e,d}(t)$ the conditional
probability of (II) given (I). Taking expectations over the arrival, we can compute
the Poisson value as in the following lemma.

\begin{restatable}[Poisson value]{lemma}{accountinglemma}\label{lem:accounting}
For any deterministic greedy policy $\pi$,
\[
 \E[\widehat M_\pi]=(2-3/e) n+\lambda\sum_{e\in E}
 \sum_{s\in\{+,-\}}\sum_{d\in\{0,1,2\}}\int_0^1h_d(t)\,P_e(t)\,Q^s_{e,d}(t)
 \,\dd t.
\]
\end{restatable}

Notice that the event (I) is independent of the policy, so we can directly
compute $P_e(t)$ from the arrival process: for $e=(u,v)$,
\[
 P_e(t)=\Prb\bigl[\Pois\bigl(2\lambda t(\deg(u)+\deg(v)-1)\bigr)=0\bigr]
 =\exp\bigl(-2\lambda t(\deg(u)+\deg(v)-1)\bigr)\geq1-4\Delta_G\lambda t.
\]

By our adjustment to the Poisson model, matching an edge type arrival does not affect the net gain of any future edge type arrival.
\begin{claim}
  Under the Poisson arrival model, an optimal policy always maximizes the net gain per edge type arrival.

\end{claim}

We also observe that when an edge type arrives, the availability of
its neighbors depends only on the allocation of previous vertex arrivals,
i.e., how the online policy chooses from $\{u^+,u^-\}$ when matching a vertex
arrival $u$. These are the central decisions that determine a policy's value.

\paragraph{The cut bonuses.}
We now specify how much net gain can be obtained from an edge arrival $e^s$ at time $t$, given the current status of its neighbors. Let $X_u(t)$ be the number of
vertex arrivals of $u$ by time $t$. Then the pair $\{u^+, u^-\}$ has both vertices free if $X_u(t)=0$,
one if $X_u(t)=1$, and none if $X_u(t)\geq2$. \Cref{tab:cases} lists all possible cases for
maximum net gains.

\begin{table}[t]
\centering
\small
\begin{tabular}{@{}lccccc@{}}
\toprule
& \multicolumn{2}{c}{Free vertices} & \multicolumn{2}{c}{Maximum net gain}
& \\
\cmidrule(lr){2-3}\cmidrule(lr){4-5}
Vertex arrivals by time $t$ & pair of $u$ & pair of $v$ & of $e^+$ & of $e^-$
& Total\\
\midrule
$X_u(t)=0$ & $u^+,u^-$ & any & \multirow{2}{*}{$h_2(t)$}
& \multirow{2}{*}{$h_2(t)$} & \multirow{2}{*}{$2h_2(t)$}\\
$X_v(t)=0$ & any & $v^+,v^-$ & & & \\
\midrule
\multirow{4}{*}{$X_u(t)=X_v(t)=1$} & $u^+$ & $v^+$ & $h_1(t)$ & $0$
& \multirow{2}{*}{\textcolor{figblue}{$\boldsymbol{h_1(t)}$}}\\
& $u^-$ & $v^-$ & $0$ & $h_1(t)$ & \\
\cmidrule(lr){2-6}
& $u^+$ & $v^-$ & $h_1(t)$ & $h_1(t)$
& \multirow{2}{*}{\textcolor{figblue}{$\boldsymbol{2h_1(t)}$}}\\
& $u^-$ & $v^+$ & $h_1(t)$ & $h_1(t)$ & \\
\midrule
\multirow{2}{*}{$X_u(t)=1$, $X_v(t)\geq2$} & $u^+$ & none & $h_1(t)$ & $0$
& \multirow{4}{*}{$h_1(t)$}\\
& $u^-$ & none & $0$ & $h_1(t)$ & \\
\multirow{2}{*}{$X_u(t)\geq2$, $X_v(t)=1$} & none & $v^+$ & $h_1(t)$ & $0$
& \\
& none & $v^-$ & $0$ & $h_1(t)$ & \\
\midrule
$X_u(t)\geq2$, $X_v(t)\geq2$ & none & none & $0$ & $0$ & $0$\\
\bottomrule
\end{tabular}
\caption{The maximum net gains of the two edge types of $e=(u,v)$ at time $t$.}
\label{tab:cases}
\end{table}

Critically, the maximum net gain of $e^s$ is determined by $(X_u(t), X_v(t))$ only,
except for the case $X_u(t)=X_v(t)=1$. In this special case, the \textit{sign} of the free vertex matters. If the free vertices of
the two pairs have different signs ($u^+, v^-$ or $u^-, v^+$), the net gain of both types is $h_1(t)$, resulting in a
total net gain of $2h_1(t)$. In contrast, when the free vertices have the same sign ($u^+, v^+$ or $u^-, v^-$), only one of $e^+$ or $e^-$ can be matched, and the total net gain is only $h_1(t)$.
Intuitively, free offline vertices with different signs create
the possibility to match either $e^+$ or $e^-$, while offline vertices with the same
sign could only be matched to one of the two edge types. This directly creates a bonus term that is associated with the cut.
Indeed, the cut-induced policy $\pi_\sigma$ leaves vertices of different signs free at the endpoints of every edge cut by $\sigma$,
and its Poisson value increases as the cut becomes larger.

\begin{restatable}[The value of a cut policy]{lemma}{cutvaluelemma}\label{lem:cutvalue}
For each cut $\sigma$ of $G$ and its induced online matching policy $\pi_\sigma$, we have
\begin{equation*}
  \Phi_G(|\delta_G(\sigma)|)-4\Delta_G\lambda^2m\leq\E[\widehat M_{\pi_\sigma}]
\leq\Phi_G(|\delta_G(\sigma)|),
\end{equation*}
where $\Phi_G$ is the affine function
\[
 \Phi_G(x)=\Bigl(2-\frac3e\Bigr)n+\Bigl(\frac1e+\frac9{e^2}\Bigr)\lambda m
 +\Bigl(\frac2e-\frac5{e^2}\Bigr)\lambda x.
\]
\end{restatable}

In $\Phi_G(x)$, the first term is the baseline from the vertex type arrivals,
the second term is the net gain of the edge type arrivals without the cut
bonuses, and the third term is the cut bonus of the $x$ edges that $\sigma$
cuts.

\subsection{Optimality of cut-induced policies}\label{sec:adaptivity}
We next show that max-cut-induced policies are near-optimal in Poisson value, once we expand the graph to be large;
no adaptive policy can substantially outperform them. Then \Cref{lem:cutvalue}
will also provide an estimate of the optimal Poisson value.

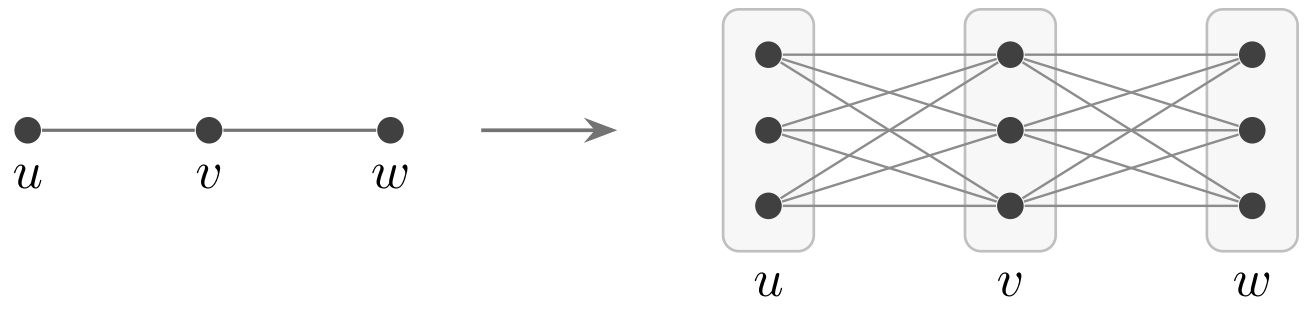
\begin{figure}[t]
\centering
\begin{tikzpicture}[font=\small]
  \foreach \n/\x in {u/0,v/1.2,w/2.4}
    \node[fig/vertex,label=below:$\n$] (g\n) at (\x,0) {};
  \draw[fig/edge] (gu)--(gv)--(gw);
  \draw[fig/maps] (3,0)--(3.9,0);
  \begin{scope}[xshift=4.9cm]
    \foreach \n/\x in {u/0,v/1.6,w/3.2} {
      \node[fig/group,minimum width=6mm,minimum height=16mm,
        label=below:$\n$] at (\x,0) {};
      \foreach \i/\y in {1/.5,2/0,3/-.5}
        \node[fig/vertex] (\n\i) at (\x,\y) {};
    }
    \foreach \i in {1,2,3}
      \foreach \j in {1,2,3}
        \draw[fig/edge,draw=black!45,line width=.45pt] (u\i)--(v\j)
          (v\i)--(w\j);
  \end{scope}
\end{tikzpicture}
\caption{A graph $G$ and its blow-up $\widetilde G$ with $B=3$.}
\label{fig:blowup}
\end{figure}

The expanded instance is $I(\widetilde G,\lambda)$, where the blow-up
$\widetilde G$ replaces every vertex of $G$ by $B$ copies
(\cref{fig:blowup}); estimating the
maximum cut size of $\widetilde G$ is as hard as for $G$. We formally define the blow-up as follows.

\begin{definition}[Blow-up]\label{def:blowup}
For an integer $B\geq1$, the \emph{blow-up} of $G$ is the graph $\widetilde G$
with vertex set $V\times\{1,\dots,B\}$, in which $(u,i)$ and $(v,j)$ are
adjacent if and only if $(u,v)\in E$. The vertices $(u,1),\dots,(u,B)$ are the
\emph{copies} of $u$.
\end{definition}

The blow-up $\widetilde G$ is a simple graph with $nB$ vertices, $mB^2$ edges,
largest degree $\Delta_{\widetilde G}=B\Delta_G$, and
max-cut value $C(\widetilde G)=B^2C(G)$. Indeed, putting every copy of $u$ on the side
$\sigma(u)$, for all $u\in V$, turns a cut $\sigma$ of $G$ into a cut of
$\widetilde G$ that is $B^2$ times larger. Conversely, giving every $u\in V$
the sign of a uniformly random copy of $u$ turns a cut of $\widetilde G$ into a
cut of $G$ that is $B^2$ times smaller in expectation, since every edge of
$\widetilde G$ joins the two chosen copies with probability $1/B^2$.

\paragraph{Bounding the gain from adaptivity.}
Unlike the cut-induced policy $\pi_\sigma$, an adaptive greedy policy may choose the signs of its free
vertices adaptively, i.e., which of $\{u^+, u^-\}$ to match first for each $u\in V$, based on the earlier arrivals. At time $t$, the vertices
with exactly one vertex arrival form a random set, which contains every vertex
independently with probability $\mathbb P[X_u(t) = 1] = te^{-t}$. On their induced subgraph, the signs of the free vertices form a cut. So at time $t$, the best
that a policy can hope for is the cut bonus of every edge in a maximum cut of
this random subgraph, while $\pi_\sigma$ earns the cut bonus of every edge in
this subgraph that $\sigma$ cuts. On the blow-up $\widetilde G$, the maximum
cut of such a random subgraph concentrates around what a fixed maximum cut
achieves.

\begin{restatable}[Random subgraphs of blow-ups]{lemma}{blowuplemma}\label{lem:blowup}
If a random set $A$ contains every vertex of $\widetilde G$ independently with
probability $p$, then
\[
 p^2C(\widetilde G)\ \leq\ \E\bigl[C(\widetilde G[A])\bigr]\ \leq\
 p^2C(\widetilde G)+mB^{3/2}.
\]
\end{restatable}

We apply \cref{lem:blowup} at every time $t$ to the vertices with one vertex
arrival by time $t$, with $p=te^{-t}$. A maximum cut $\sigma$ of $\widetilde G$
attains the lower bound, so choosing the signs adaptively gains only
$O(\lambda mB^{3/2})$ in Poisson value over $\pi_\sigma$.

\begin{restatable}[Adaptivity gap]{lemma}{gaplemma}\label{lem:gap}
Let $\sigma$ be a maximum cut of $\widetilde G$. On the instance
$I(\widetilde G,\lambda)$,
\[
 \E[\widehat M_{\pi_\sigma}]\ \leq\ \widehat\OPT\ \leq\
 \E[\widehat M_{\pi_\sigma}]+\lambda mB^{3/2}+4\Delta_G\lambda^2mB^3.
\]
\end{restatable}

The proof (\cref{app:gap}) shows
$\widehat\OPT\leq\Phi_{\widetilde G}(C(\widetilde G))+\lambda mB^{3/2}$ and then
applies \cref{lem:cutvalue} to $\widetilde G$, whose blocking error is
$4\Delta_{\widetilde G}\lambda^2mB^2=4\Delta_G\lambda^2mB^3$.

\paragraph{The reduction.}
For a cubic graph $G$, by combining \cref{cor:transfer}, \cref{lem:cutvalue}, and \cref{lem:gap}, we have
\[
 \bigl|\OPT(I(\widetilde G,\lambda))-\Phi_{\widetilde G}\bigl(B^2C(G)\bigr)\bigr|
 \leq\lambda mB^2\Bigl(24\lambda B+\frac1{\sqrt B}\Bigr)
 +\sqrt{nB+2\lambda mB^2}+1.
\]
As $\Phi_{\widetilde G}(B^2x)$ is affine in $x$ with slope
$(\frac2e-\frac5{e^2})\lambda B^2$, the online optimum determines $C(G)$ up to
an error of $O(\lambda B+1/\sqrt B)\,m$ for large $n$. We take $B$ large and
$\lambda B$ small, so that this error is a small fraction of $C(G)\geq m/2$.
Then an approximation of the online optimum within a small enough constant
factor gives an approximation of $C(G)$ within the factor $1\pm\eta_0$ of
\cref{thm:maxcut}.

\paragraph{Randomized algorithms.}
By a similar reduction, we can also show the hardness result for polynomial-time, randomized online policies.

\begin{restatable}{theorem}{randthm}\label{thm:randomized}
For the constant $\varepsilon_0$ and the instances of \cref{thm:main}, unless
$\NP\subseteq\BPP$, no randomized polynomial-time algorithm approximates the
online optimum within a factor $1\pm\varepsilon_0$ with probability at least
$2/3$, and no randomized polynomial-time algorithm outputs, on every instance
$I$, a polynomial-time online policy with expected value at least
$(1-\varepsilon_0/2)\OPT(I)$.
\end{restatable}

\section{$\SharpP$-Hardness of Prophet Secretary}\label{sec:secretary}

In this section, we prove the $\SharpP$-hardness of computing an optimal policy
for prophet secretary.

\begin{restatable}{theorem}{decisionsthm}\label{thm:decisions}
For prophet secretary, where the values arrive in a uniformly random order,
computing an optimal policy is $\SharpP$-hard, even when every value is $0$ or
$1$, except for two deterministic values in $(0,1)$.
\end{restatable}

We reduce from \CountSS, the problem of finding the number of subsets whose elements sum up to a target value. An instance of the \CountSS{} problem consists of positive
integers $a_1,\dots,a_n$ and a target $b\in(0,A]$, where
$A=\sum_{i=1}^na_i$. Each solution is a subset $S\subseteq[n]$ with
$\sum_{i\in S}a_i=b$, and the question is to count all solutions.
The problem is known to be $\SharpP$-complete~\cite{FH09}. Since multiplying all numbers by a common positive factor does not change the solutions, we can always scale the problem such that $a_1, \dots, a_n$ and $b$ are rationals that are multiples of a small rational $\varepsilon>0$. By choosing small $\varepsilon$, we can always let $2A^2 <\varepsilon$.

We only consider instances of \CountSS{} where all solutions share the same cardinality $k\in \mathbb{N}$. This does not affect the $\SharpP$-hardness of the problem.

\paragraph{From values to decisions.}
We will first show that computing the online optimal \textit{value} is hard, when only
one value is deterministic and the optimal policy itself is easy to find.

\begin{restatable}[Hardness of the online optimum]{lemma}{optlemma}\label{lem:opt}
For prophet secretary, computing the online optimum is $\SharpP$-hard, even
when every value is $0$ or $1$, except for one deterministic value in $(0,1)$.
\end{restatable}

We can then construct a hard instance for finding the optimal policy by adding one online vertex with a deterministic value.
Given an instance $I$ as in \Cref{lem:opt} and a new online vertex with deterministic value $q$, with probability $1/(T+1)$ the new vertex arrives first, and in this case the optimal policy needs to decide if $q \geq \OPT(I)$. If the optimal policy can be constructed for each $q \in (0,1)$, we can then efficiently compute $\OPT(I)$ by a binary search over $q$ (\cref{app:decisions}). In the rest of this section, we will focus on \Cref{lem:opt}.

\paragraph{Prophet secretary instance from \CountSS.}
We now turn a scaled instance of \CountSS{} into prophet secretary instances whose online optimum values characterize the number $N$ of solutions.
For $x\in(0,1)$, we consider the following prophet secretary instance $I(x)$ with $T = n+1$ online vertices:
\begin{itemize}[leftmargin=*,itemsep=1pt]
\item \textit{Random 0-1 values}: For each $i\in[n]$, we construct an online
vertex $v_i$ that has value $1$ with probability $a_i$, and value $0$
otherwise.
\item \textit{Deterministic value}: We construct one online vertex $v_x$ that
has the deterministic value $x$.
\end{itemize}
We identify each set $S\subseteq[n]$ with the set of online vertices $v_i$, $i\in S$.
As we scale down $a_1, \dots, a_n$ and $b$ uniformly, each online vertex $v_i$ has value $1$ only with small probability, and the probability for some online vertex of
a set $S\subseteq[n]$ to have value $1$ is approximately $\sum_{i\in S}a_i$, i.e., the sum of its elements.

Finding the optimal policy for $I(x)$ is simple. Since all values are in $[0,1]$, an optimal policy will always accept a value $1$ and reject a value $0$, so the only real decision is whether to accept $x$.
By backward induction, the optimal policy accepts value $x$ when it is greater than the maximum expected future gain, i.e., the probability that the future arrivals contain a value $1$:
\[
 r_S=1-\prod_{i\in S}(1-a_i)\approx\sum_{i\in S}a_i,
\]
where $S \subseteq [n]$ is the set of arrivals after $v_x$.

Without $v_x$, the optimal policy accepts the first value $1$, so the online
optimum is $r_{[n]}$. Adding $v_x$ changes the expected reward only when all
values before $v_x$ are $0$, from $r_S$ to $\max\{x,r_S\}$. Hence, for any
$x\in(0,1)$, $\OPT(I(x))$ is a piecewise linear function of $x$,
\begin{equation}\label{eq:value}
 \OPT(I(x))=r_{[n]}+\sum_{S\subseteq[n]}P_S\max\{x-r_S,0\},
\end{equation}
where $P_S$ is the probability that $S$ is exactly the set of online vertices that arrive after $v_x$ and all values before $v_x$ are $0$, i.e.,
\[
 P_S=\frac1{(n+1)\binom n{|S|}}\prod_{i\in[n]\setminus S}(1-a_i).
\]

While the optimal policy for $I(x)$ is simple, the sum in \eqref{eq:value}
has exponentially many terms, and computing the value of $\OPT(I(x))$ is hard.

\paragraph{Breakpoints of the online optimum.}
$\OPT(I(x))$ is convex and piecewise linear in $x$, with a
breakpoint at $r_S$ for every set $S\subseteq[n]$. Since
$r_S\approx\sum_{i\in S}a_i$, the breakpoints $r_{S_1}$ and $r_{S_2}$ are close
together when $S_1$ and $S_2$ have the same sum of elements. Meanwhile, we argue that
breakpoints of sets with different sums are far apart.
\begin{align*}
 \sum_{i\in S}a_i\ \geq\ r_S=1-\prod_{i\in S}(1-a_i)
 &\ \geq\ 1-\exp\Bigl(-\sum_{i\in S}a_i\Bigr)
 \ \geq\ \sum_{i\in S}a_i-\frac12\Bigl(\sum_{i\in S}a_i\Bigr)^2\\
 &\ \geq\ \sum_{i\in S}a_i-\frac{A^2}2
 \ >\ \sum_{i\in S}a_i-\frac\varepsilon4, \ \forall S\subseteq[n].
\end{align*}
Therefore, $r_S$ lies in the interval $(b-\varepsilon/4,b+\varepsilon/4)$ around $b$ if
and only if $\sum_{i\in S}a_i=b$, i.e., $S$ is a solution to be counted.

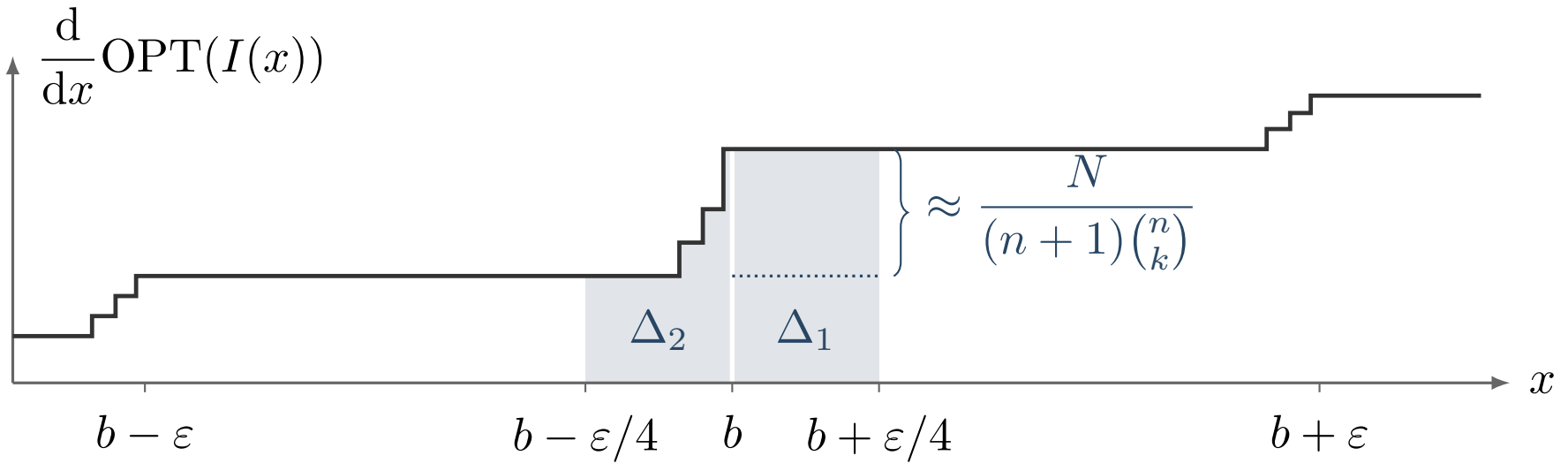
\begin{figure}[t]
\centering
\begin{tikzpicture}[x=1.1cm,y=1cm,font=\small]
  \fill[figblue!14] (-1,0) |- (-.36,.8) |- (-.2,1.05) |- (-.06,1.3)
    |- (0,1.75) -- (0,0) -- cycle;
  \fill[figblue!14] (0,0) rectangle (1,1.75);
  \draw[white,line width=1pt] (0,0)--(0,1.75);
  \draw[figblue,densely dotted,line width=.6pt] (0,.8)--(1,.8);
  \node[text=figblue] at (-.5,.4) {$\Delta_2$};
  \node[text=figblue] at (.5,.4) {$\Delta_1$};
  \draw[fig/axis] (-4.9,0)--(5.3,0) node[right] {$x$};
  \draw[fig/axis] (-4.9,0)--(-4.9,2.45)
    node[right=1pt] {$\displaystyle\frac{\dd}{\dd x}\OPT(I(x))$};
  \foreach \p/\l in {-4/{b-\varepsilon},-1/{b-\varepsilon/4},0/b,
      1/{b+\varepsilon/4},4/{b+\varepsilon}} {
    \draw[black!60] (\p,0)--(\p,-.07);
    \node[below=1pt] at (\p,-.07) {$\l$};
  }
  \draw[black!80,line width=.9pt] (-4.9,.35) -| (-4.36,.5) -| (-4.2,.65)
    -| (-4.06,.8) -| (-.36,1.05) -| (-.2,1.3) -| (-.06,1.75) -| (3.64,1.9)
    -| (3.8,2.02) -| (3.94,2.15) -- (5.1,2.15);
  \draw[figblue,decorate,decoration={brace,amplitude=3pt}] (1.1,1.75)
    --(1.1,.8) node[midway,right=3pt]
    {$\displaystyle\approx\frac N{(n+1)\binom nk}$};
\end{tikzpicture}
\caption{The slope of $\OPT(I(x))$ as a function of $x$.}
\label{fig:slope}
\end{figure}

To count the solutions, we investigate the slope of $\OPT(I(x))$.
At each $x$ that is not a breakpoint, we have
\[
 \frac{\dd}{\dd x}\OPT(I(x))=\sum_{S\subseteq[n]:\ r_S<x}P_S.
\]
This slope is a step function of $x$ that increases by $P_S$ at $r_S$
(\cref{fig:slope}).
On the interval $(b-\varepsilon/4,b+\varepsilon/4)$, the slope increases only at the
breakpoints of the solutions since there are no other breakpoints in the interval. Recall that all solutions have $k$ elements, so the increment is
\[
 \sum_{S\subseteq[n]:\ \sum_{i\in S}a_i=b}P_S
 =\frac1{(n+1)\binom nk}\sum_{S\subseteq[n]:\ \sum_{i\in S}a_i=b}\
 \prod_{i\in[n]\setminus S}(1-a_i)
 \approx\frac N{(n+1)\binom nk}.
\]

If $\OPT(I(x))$ can be computed efficiently, we can then estimate this increment in the slope of $\OPT(I(x))$. We compare the increases of
$\OPT(I(x))$ over the two halves of the interval:
\begin{align*}
 \Delta_1=\OPT(I(b+\varepsilon/4))-\OPT(I(b)),\quad \Delta_2=\OPT(I(b))-\OPT(I(b-\varepsilon/4)).
\end{align*}
As the breakpoints of the solutions are close to $b$, $\Delta_1-\Delta_2$ can be used to estimate the increment of the slope, which gives the
following estimate of $N$.

\begin{restatable}[Counting]{lemma}{countinglemma}\label{lem:counting}
We have
\[
 \Bigl(1-\frac{3A^2}\varepsilon\Bigr)N\ \leq\ \frac{4(n+1)}\varepsilon\cdot \binom nk \cdot
 (\Delta_1-\Delta_2)\ \leq\ N.
\]
\end{restatable}

\section*{AI Usage Disclosure}
The main hardness proofs were obtained from extensive interactions with GPT 6 Astra and Gemini 4 Argon. The main body of the paper is written by the authors, with wording and image illustration polished by LLMs. The authors take full responsibility for all the content.

\bibliographystyle{plainnat}
\bibliography{references}

\begin{thebibliography}{54}
\providecommand{\natexlab}[1]{#1}
\providecommand{\url}[1]{\texttt{#1}}
\expandafter\ifx\csname urlstyle\endcsname\relax
  \providecommand{\doi}[1]{doi: #1}\else
  \providecommand{\doi}{doi: \begingroup \urlstyle{rm}\Url}\fi

\bibitem[Agrawal et~al.(2019)Agrawal, Sethuraman, and Zhang]{ASZ20}
Shipra Agrawal, Jay Sethuraman, and Xingyu Zhang.
\newblock On optimal ordering in the optimal stopping problem.
\newblock 2019.

\bibitem[Alimonti and Kann(2000)]{AK00}
Paola Alimonti and Viggo Kann.
\newblock Some apx-completeness results for cubic graphs.
\newblock \emph{Theoretical Computer Science}, 237\penalty0 (1-2):\penalty0
  123--134, 2000.

\bibitem[Anari et~al.(2019)Anari, Niazadeh, Saberi, and Shameli]{ANSS19}
Nima Anari, Rad Niazadeh, Amin Saberi, and Ali Shameli.
\newblock Nearly optimal pricing algorithms for production constrained and
  laminar bayesian selection.
\newblock In \emph{Proceedings of the 2019 ACM Conference on Economics and
  Computation}, pages 91--92, 2019.

\bibitem[Azar et~al.(2018)Azar, Chiplunkar, and Kaplan]{ACK18}
Yossi Azar, Ashish Chiplunkar, and Haim Kaplan.
\newblock Prophet secretary: Surpassing the 1-1/e barrier.
\newblock In \emph{Proceedings of the 2018 ACM Conference on Economics and
  Computation}, pages 303--318, 2018.

\bibitem[Beyhaghi and Cai(2023)]{BC23}
Hedyeh Beyhaghi and Linda Cai.
\newblock Pandora’s problem with nonobligatory inspection: Optimal structure
  and a ptas.
\newblock In \emph{Proceedings of the 55th Annual ACM Symposium on Theory of
  Computing}, pages 803--816, 2023.

\bibitem[Braun et~al.(2024)Braun, Kesselheim, Pollner, and Saberi]{BKPS24}
Alexander Braun, Thomas Kesselheim, Tristan Pollner, and Amin Saberi.
\newblock Approximating optimum online for capacitated resource allocation.
\newblock 2024.

\bibitem[Braverman et~al.(2022)Braverman, Derakhshan, and Molina~Lovett]{BDL22}
Mark Braverman, Mahsa Derakhshan, and Antonio Molina~Lovett.
\newblock Max-weight online stochastic matching: Improved approximations
  against the online benchmark.
\newblock In \emph{Proceedings of the 23rd ACM Conference on Economics and
  Computation}, pages 967--985, 2022.

\bibitem[Braverman et~al.(2025)Braverman, Derakhshan, Pollner, Saberi, and
  Wajc]{BDPSW25}
Mark Braverman, Mahsa Derakhshan, Tristan Pollner, Amin Saberi, and David Wajc.
\newblock New philosopher inequalities for online bayesian matching, via
  pivotal sampling.
\newblock In \emph{Proceedings of the 2025 Annual ACM-SIAM Symposium on
  Discrete Algorithms (SODA)}, pages 3029--3068. SIAM, 2025.

\bibitem[Bubna and Chiplunkar(2023)]{BC23ec}
Archit Bubna and Ashish Chiplunkar.
\newblock Prophet inequality: Order selection beats random order.
\newblock In \emph{Proceedings of the 24th ACM Conference on Economics and
  Computation}, pages 302--336, 2023.

\bibitem[Chen et~al.(2025)Chen, Huang, Li, and Tang]{CHLT26}
Ziyun Chen, Zhiyi Huang, Dongchen Li, and Zhihao~Gavin Tang.
\newblock Prophet secretary and matching: the significance of the largest item.
\newblock ACM, New York, NY, 2025.

\bibitem[Chierichetti et~al.(2025)Chierichetti, Giacchini, Panconesi, and
  Vattani]{CGPV25}
Flavio Chierichetti, Mirko Giacchini, Alessandro Panconesi, and Andrea Vattani.
\newblock A new impossibility result for online bipartite matching problems.
\newblock 2025.

\bibitem[Correa et~al.(2019)Correa, Foncea, Hoeksma, Oosterwijk, and
  Vredeveld]{CFHOV18}
Jose Correa, Patricio Foncea, Ruben Hoeksma, Tim Oosterwijk, and Tjark
  Vredeveld.
\newblock Recent developments in prophet inequalities.
\newblock \emph{ACM SIGecom Exchanges}, 17\penalty0 (1):\penalty0 61--70, 2019.

\bibitem[Correa et~al.(2021{\natexlab{a}})Correa, Foncea, Hoeksma, Oosterwijk,
  and Vredeveld]{CFHOV21}
Jos{\'e} Correa, Patricio Foncea, Ruben Hoeksma, Tim Oosterwijk, and Tjark
  Vredeveld.
\newblock Posted price mechanisms and optimal threshold strategies for random
  arrivals.
\newblock \emph{Mathematics of operations research}, 46\penalty0 (4):\penalty0
  1452--1478, 2021{\natexlab{a}}.

\bibitem[Correa et~al.(2021{\natexlab{b}})Correa, Saona, and Ziliotto]{CSZ21}
Jose Correa, Raimundo Saona, and Bruno Ziliotto.
\newblock Prophet secretary through blind strategies: J. correa et al.
\newblock \emph{Mathematical Programming}, 190\penalty0 (1):\penalty0 483--521,
  2021{\natexlab{b}}.

\bibitem[DeHaan and Pashkovich(2024)]{DP24}
Ian DeHaan and Kanstantsin Pashkovich.
\newblock Matroid bayesian online selection.
\newblock In \emph{International Symposium on Algorithmic Game Theory}, pages
  405--422. Springer, 2024.

\bibitem[Doval(2018)]{Dov18}
Laura Doval.
\newblock Whether or not to open pandora's box.
\newblock \emph{Journal of Economic Theory}, 175:\penalty0 127--158, 2018.

\bibitem[Duetting et~al.(2026)Duetting, Feldman, and Molina]{DFM26}
Paul Duetting, Michal Feldman, and Mathieu Molina.
\newblock Optimal online stochastic matching, 2026.

\bibitem[D{\"u}tting et~al.(2023)D{\"u}tting, Gergatsouli, Rezvan, Teng, and
  Tsigonias-Dimitriadis]{DGRTT23}
Paul D{\"u}tting, Evangelia Gergatsouli, Rojin Rezvan, Yifeng Teng, and
  Alexandros Tsigonias-Dimitriadis.
\newblock Prophet secretary against the online optimal.
\newblock In \emph{Proceedings of the 24th ACM Conference on Economics and
  Computation}, pages 561--581, 2023.

\bibitem[Ehsani et~al.(2018)Ehsani, Hajiaghayi, Kesselheim, and Singla]{EHKS18}
Soheil Ehsani, MohammadTaghi Hajiaghayi, Thomas Kesselheim, and Sahil Singla.
\newblock Prophet secretary for combinatorial auctions and matroids.
\newblock In \emph{Proceedings of the twenty-ninth annual acm-siam symposium on
  discrete algorithms}, pages 700--714. SIAM, 2018.

\bibitem[Esfandiari et~al.(2017)Esfandiari, Hajiaghayi, Liaghat, and
  Monemizadeh]{EHLM17}
Hossein Esfandiari, MohammadTaghi Hajiaghayi, Vahid Liaghat, and Morteza
  Monemizadeh.
\newblock Prophet secretary.
\newblock \emph{SIAM Journal on Discrete Mathematics}, 31\penalty0
  (3):\penalty0 1685--1701, 2017.

\bibitem[Faliszewski and Hemaspaandra(2009)]{FH09}
Piotr Faliszewski and Lane Hemaspaandra.
\newblock The complexity of power-index comparison.
\newblock \emph{Theoretical Computer Science}, 410\penalty0 (1):\penalty0
  101--107, 2009.

\bibitem[Feldman et~al.(2009)Feldman, Mehta, Mirrokni, and
  Muthukrishnan]{FMMM09}
Jon Feldman, Aranyak Mehta, Vahab Mirrokni, and Shan Muthukrishnan.
\newblock Online stochastic matching: Beating 1-1/e.
\newblock In \emph{2009 50th Annual IEEE Symposium on Foundations of Computer
  Science}, pages 117--126. IEEE, 2009.

\bibitem[Feldman et~al.(2014)Feldman, Gravin, and Lucier]{FGL15}
Michal Feldman, Nick Gravin, and Brendan Lucier.
\newblock Combinatorial auctions via posted prices.
\newblock In \emph{Proceedings of the twenty-sixth annual ACM-SIAM symposium on
  Discrete algorithms}, pages 123--135. SIAM, 2014.

\bibitem[Feng et~al.(2026)Feng, Jin, Wu, and Xiang]{FJWX26}
Yilong Feng, Yaonan Jin, Xiaowei Wu, and Xinyue Xiang.
\newblock Mean-only prophet inequalities, 2026.

\bibitem[Fu et~al.(2023)Fu, Li, and Liu]{FLL23}
Hu~Fu, Jiawei Li, and Daogao Liu.
\newblock Pandora box problem with nonobligatory inspection: Hardness and
  approximation scheme.
\newblock In \emph{Proceedings of the 55th Annual ACM Symposium on Theory of
  Computing}, pages 789--802, 2023.

\bibitem[Giambartolomei et~al.(2023)Giambartolomei, Mallmann-Trenn, and
  Saona]{GMS23}
Giordano Giambartolomei, Frederik Mallmann-Trenn, and Raimundo Saona.
\newblock Prophet inequalities: Separating random order from order selection,
  2023.

\bibitem[Goel and Mehta(2008)]{GM08}
Gagan Goel and Aranyak Mehta.
\newblock Online budgeted matching in random input models with applications to
  adwords.
\newblock In \emph{SODA}, volume~8, pages 982--991, 2008.

\bibitem[Harb(2025)]{Harb25}
Elfarouk Harb.
\newblock New prophet inequalities via poissonization and sharding.
\newblock In \emph{Proceedings of the 2025 Annual ACM-SIAM Symposium on
  Discrete Algorithms (SODA)}, pages 1222--1269. SIAM, 2025.

\bibitem[Hill and Kertz(1982)]{HK82}
Theodore~P Hill and Robert~P Kertz.
\newblock Comparisons of stop rule and supremum expectations of iid random
  variables.
\newblock \emph{The Annals of Probability}, pages 336--345, 1982.

\bibitem[Huang and Shu(2021)]{HS21}
Zhiyi Huang and Xinkai Shu.
\newblock Online stochastic matching, poisson arrivals, and the natural linear
  program.
\newblock In \emph{Proceedings of the 53rd Annual ACM SIGACT Symposium on
  Theory of Computing}, pages 682--693, 2021.

\bibitem[Huang et~al.(2022)Huang, Shu, and Yan]{HSY22}
Zhiyi Huang, Xinkai Shu, and Shuyi Yan.
\newblock The power of multiple choices in online stochastic matching.
\newblock In \emph{Proceedings of the 54th Annual ACM SIGACT Symposium on
  Theory of Computing}, pages 91--103, 2022.

\bibitem[Huang et~al.(2024)Huang, Tang, and Wajc]{HTW24}
Zhiyi Huang, Zhihao~Gavin Tang, and David Wajc.
\newblock Online matching: A brief survey.
\newblock \emph{ACM SIGecom Exchanges}, 22\penalty0 (1):\penalty0 135--158,
  2024.

\bibitem[Karp et~al.(1990)Karp, Vazirani, and Vazirani]{KVV90}
Richard~M Karp, Umesh~V Vazirani, and Vijay~V Vazirani.
\newblock An optimal algorithm for on-line bipartite matching.
\newblock In \emph{Proceedings of the twenty-second annual ACM symposium on
  Theory of computing}, pages 352--358, 1990.

\bibitem[Kertz(1986)]{Ker86}
Robert~P Kertz.
\newblock Stop rule and supremum expectations of iid random variables: a
  complete comparison by conjugate duality.
\newblock \emph{Journal of multivariate analysis}, 19\penalty0 (1):\penalty0
  88--112, 1986.

\bibitem[Kesselheim et~al.(2013)Kesselheim, Radke, T{\"o}nnis, and
  V{\"o}cking]{KRTV13}
Thomas Kesselheim, Klaus Radke, Andreas T{\"o}nnis, and Berthold V{\"o}cking.
\newblock An optimal online algorithm for weighted bipartite matching and
  extensions to combinatorial auctions.
\newblock In \emph{European symposium on algorithms}, pages 589--600. Springer,
  2013.

\bibitem[Krengel and Sucheston(1978)]{KS78}
Ulrich Krengel and Louis Sucheston.
\newblock On semiamarts, amarts, and processes with finite value.
\newblock volume~4, pages 1--2. 1978.

\bibitem[Liu et~al.(2020)Liu, Leme, P{\'a}l, Schneider, and Sivan]{LLPSS21}
Allen Liu, Renato~Paes Leme, Martin P{\'a}l, Jon Schneider, and Balasubramanian
  Sivan.
\newblock Variable decomposition for prophet inequalities and optimal ordering.
\newblock 2020.

\bibitem[Lucier(2017)]{Luc17}
Brendan Lucier.
\newblock An economic view of prophet inequalities.
\newblock \emph{ACM SIGecom Exchanges}, 16\penalty0 (1):\penalty0 24--47, 2017.

\bibitem[Mahdian and Yan(2011)]{MY11}
Mohammad Mahdian and Qiqi Yan.
\newblock Online bipartite matching with random arrivals: an approach based on
  strongly factor-revealing lps.
\newblock In \emph{Proceedings of the forty-third annual ACM symposium on
  Theory of computing}, pages 597--606, 2011.

\bibitem[Manshadi et~al.(2012)Manshadi, Gharan, and Saberi]{MGS12}
Vahideh~H Manshadi, Shayan~Oveis Gharan, and Amin Saberi.
\newblock Online stochastic matching: Online actions based on offline
  statistics.
\newblock \emph{Mathematics of Operations Research}, 37\penalty0 (4):\penalty0
  559--573, 2012.

\bibitem[Mehta(2010)]{Meh13}
Aranyak Mehta.
\newblock Online matching and ad allocation.
\newblock \emph{Foundations and Trends{\textregistered} in Theoretical Computer
  Science}, 8\penalty0 (4):\penalty0 265--368, 2010.

\bibitem[Naor et~al.(2025)Naor, Srinivasan, and Wajc]{NSW23}
Joseph Naor, Aravind Srinivasan, and David Wajc.
\newblock Online dependent rounding schemes for bipartite matchings, with
  applications.
\newblock In \emph{Proceedings of the 2025 Annual ACM-SIAM Symposium on
  Discrete Algorithms (SODA)}, pages 3116--3154. SIAM, 2025.

\bibitem[Oliveira et~al.(2026)Oliveira, Singh, and Singla]{OSS26}
Thiago Oliveira, Mohit Singh, and Sahil Singla.
\newblock Philosopher and prophet inequalities for divisible items, 2026.

\bibitem[Papadimitriou et~al.(2024)Papadimitriou, Pollner, Saberi, and
  Wajc]{PPSW21}
Christos Papadimitriou, Tristan Pollner, Amin Saberi, and David Wajc.
\newblock Online stochastic max-weight bipartite matching: Beyond prophet
  inequalities.
\newblock volume~49, pages 1607--1628. INFORMS, 2024.

\bibitem[Papadimitriou(1985)]{Pap85}
Christos~H Papadimitriou.
\newblock Games against nature.
\newblock \emph{Journal of Computer and System Sciences}, 31\penalty0
  (2):\penalty0 288--301, 1985.

\bibitem[Peng and Tang(2025)]{PT25}
Bo~Peng and Zhihao~Gavin Tang.
\newblock Revisiting ranking for online bipartite matching with random
  arrivals: the primal-dual analysis.
\newblock In \emph{Proceedings of the 26th ACM Conference on Economics and
  Computation}, pages 820--835, 2025.

\bibitem[Saberi and Wajc(2021)]{SW21}
Amin Saberi and David Wajc.
\newblock The greedy algorithm is not optimal for on-line edge coloring.
\newblock In \emph{48th International Colloquium on Automata, Languages, and
  Programming (ICALP 2021)}, pages 109--1. Schloss Dagstuhl--Leibniz-Zentrum
  f{\"u}r Informatik, 2021.

\bibitem[Samuel-Cahn(1984)]{SC84}
Ester Samuel-Cahn.
\newblock Comparison of threshold stop rules and maximum for independent
  nonnegative random variables.
\newblock \emph{the Annals of Probability}, pages 1213--1216, 1984.

\bibitem[Segev and Singla(2021)]{SS25}
Danny Segev and Sahil Singla.
\newblock Efficient approximation schemes for stochastic probing and prophet
  problems.
\newblock In \emph{Proceedings of the 22nd ACM Conference on Economics and
  Computation}, pages 793--794, 2021.

\bibitem[Sun et~al.(2025)Sun, Tang, and Wang]{STW25}
Enze Sun, Zhihao~Gavin Tang, and Yifan Wang.
\newblock Online stochastic matching with unknown arrival order: Beating 0.5
  against the online optimum.
\newblock In \emph{Proceedings of the 57th Annual ACM Symposium on Theory of
  Computing}, pages 1430--1441, 2025.

\bibitem[Sun et~al.(2026)Sun, Tang, and Wang]{STW26}
Enze Sun, Zhihao~Gavin Tang, and Yifan Wang.
\newblock Combinatorial philosopher inequalities.
\newblock In \emph{Proceedings of the 2026 Annual ACM-SIAM Symposium on
  Discrete Algorithms (SODA)}, pages 3872--3904. SIAM, 2026.

\bibitem[Vuong et~al.(2025)Vuong, Kleinberg, Prasad, Xiao, and Xu]{DCKPXX25}
Ramiro~N Vuong, Robert Kleinberg, Aditya Prasad, Eric Xiao, and Haifeng Xu.
\newblock The keychain problem: On minimizing the opportunity cost of
  uncertainty.
\newblock 2025.

\bibitem[Weitzman(1978)]{Wei79}
Martin Weitzman.
\newblock Optimal search for the best alternative.
\newblock 78\penalty0 (8), 1978.

\bibitem[Yan(2025)]{Yan26}
Shuyi Yan.
\newblock Edge-weighted online stochastic matching under jaillet-lu lp.
\newblock 2025.

\end{thebibliography}

\clearpage
\appendix
\crefalias{section}{appendix}
\section{Related work}\label{sec:related}

\paragraph{Online matching against the offline optimum.}
Much of the work on online bipartite matching has focused on competing with
the offline optimum, initiated by~\cite{KVV90}.
As the literature is too large, we only discuss random-order or stochastic settings and their best results here\footnote{With our best effort: we are aware that most competitive ratios for matching and prophet inequality settings cited below have already been improved recently.}; we refer the readers to \cite{Meh13,HTW24} for surveys of the online matching literature.
Many papers studied whether random arrivals allow better guarantees,
either by presenting unknown online vertices in random order~\cite{GM08}
or by drawing requests independently from a known distribution~\cite{FMMM09}.
In the random-order model, the best known competitive ratios against the offline optimum are $0.696$ for unweighted
matching~\cite{MY11}, $0.6862$ for offline vertex-weighted
matching~\cite{PT25}, and $1/e$ for edge-weighted matching~\cite{KRTV13} (which is tight for the secretary problem).
For known i.i.d.\ arrivals, the best known polynomial-time competitive ratios are $0.716$
for unweighted and offline vertex-weighted matching~\cite{HSY22} and
$0.662$ for edge-weighted matching~\cite{Yan26}. For unweighted known-i.i.d.\ matching, the upper bound of
$1-e/e^e\approx 0.8206$~\cite{CGPV25} was recently matched asymptotically
by an exponential-time online policy~\cite{DFM26}.
With known non-identical distributions, edge-weighted matching has a tight
ratio of $1/2$ in fixed order~\cite{KS78,FGL15}. For random order, the $1-1/e$ competitive ratio of~\cite{EHKS18} has been improved to $0.641$~\cite{CHLT26}.

\paragraph{Approximating the online optimum.}
When arrival distributions are known, the optimal online policy is a natural
benchmark, as already suggested in~\cite{MGS12}.
For matching, \cite{PPSW21} first proved that there exists a polynomial-time
algorithm with a $0.51$-approximation for fixed-order arrivals.
Subsequent rounding algorithms improved this guarantee~\cite{SW21,BDL22,NSW23},
with the best current ratio for edge-weighted matching being $0.678$~\cite{BDPSW25}.
Such approximation guarantees, also known as
\emph{philosopher inequalities}, have also been obtained
for capacitated resource allocation~\cite{BKPS24}, matching with an unknown
arrival order~\cite{STW25}, submodular combinatorial
allocation~\cite{STW26}, and divisible items~\cite{OSS26}.
Stronger guarantees are possible for laminar Bayesian selection:
PTAS algorithms are known for bounded-depth constraints~\cite{ANSS19} and for
arbitrary depth under a left-to-right arrival order~\cite{DP24}.

\paragraph{Computational hardness of the online optimum.}
For fixed-order Bayesian matching, approximating the online optimum
within some constant factor is $\PSPACE$-hard~\cite{PPSW21}, even in
unweighted graphs~\cite{BDPSW25}.
For weighted matching with correlated arrivals, approximating the online optimum is $\APX$-hard when the joint distribution is given as an explicit list of scenarios~\cite{DCKPXX25}.\footnote{\cite{MGS12} remark
\emph{without proof} that computing an optimal policy for online stochastic
matching is $\NP$-hard. As later work with a common author~\cite{PPSW21} did
not credit that paper for proving $\NP$-hardness, we also do not credit it for
the hardness result.}
Beyond matching, constant-factor approximation is $\PSPACE$-hard for
Bayesian selection under graphic matroid constraints~\cite{DP24}.
Exact policy computation is $\SharpP$-hard even for a single divisible
item~\cite{OSS26}.
These hardness constructions use fixed arrival orders and do not
establish hardness under i.i.d.\ arrivals or a random arrival order.

\paragraph{Prophet secretary and single-item optimal stopping.}
For independent arrivals and a single offline vertex, the matching problem with known distributions becomes the single-item optimal stopping problem. The tight competitive ratio against the offline
optimum is $1/2$ in fixed order~\cite{KS78,SC84} and approximately $0.745$
for known i.i.d.\ arrivals~\cite{HK82,Ker86,CFHOV21};
see~\cite{Luc17,CFHOV18} for surveys of the prophet inequality problems.
With a uniformly random arrival order, this becomes the prophet secretary
problem~\cite{EHLM17}.
Algorithms beating $1-1/e$ against the offline optimum have been developed in~\cite{ACK18,CSZ21,Harb25,CHLT26},
and the best known competitive ratio is now $0.6902$~\cite{FJWX26}.
The best known upper bound is $0.7235$~\cite{GMS23}, below the $0.7258$
guarantee possible when the policy chooses the arrival order~\cite{BC23ec}.

The optimal online policy in fixed order can be computed in polynomial
time via backward induction.
For prophet secretary, \cite{DGRTT23} gave a PTAS for the online optimum.
Our $\SharpP$-hardness result shows that exact policy computation is hard
even though arbitrarily accurate approximation is possible.
In the related free-order prophet problem, finding an optimal arrival order
is $\NP$-hard~\cite{ASZ20} but admits an EPTAS~\cite{LLPSS21,SS25}. Another optimal stopping problem with similar approximation and hardness results is the Pandora's box~\cite{Wei79} with nonobligatory inspection~\cite{Dov18} problem: computing its optimal policy is
$\NP$-hard but admits a PTAS~\cite{BC23,FLL23}.

\section{Proofs for Section~\ref{sec:matching}}\label{app:matching}

For a deterministic policy $\pi$ and $\ell\geq0$, let $\widehat M_\pi(\ell)$
be the number of matches of $\pi$ on the first $\ell$ arrivals of the
adjusted Poisson model, so that $\widehat M_\pi=\widehat M_\pi(K(1))$.

\subsection{Poisson arrivals}\label{app:poisson}

As in \cref{sec:estimate}, the arrivals of the different types form
independent Poisson processes, where type $y$ has rate $\lambda_y$. In
particular, for $t\in[0,1]$, the numbers $X_u(t)$, $u\in V$, are independent
and have distribution $\Pois(t)$.

\begin{lemma}\label{lem:poisson}
For $\ell\geq0$, let $F_\ell$ map an $\ell$-tuple of types and a time in
$[0,1]$ to $[0,1]$. Then for every type $y$,
\[
 \E\Biggl[\sum_{k\geq1:\,\tau_k\leq1}\ind{y_k=y}\,
 F_{k-1}(y_1,\dots,y_{k-1},\tau_k)\Biggr]
 =\lambda_y\int_0^1\E\bigl[F_{K(t)}(y_1,\dots,y_{K(t)},t)\bigr]\,\dd t.
\]
\end{lemma}

\begin{proof}
Fix $k\geq1$. The point $\tau_k$ has the density
$s\mapsto R^ks^{k-1}e^{-Rs}/(k-1)!=R\,\Prb(K(s)=k-1)$. We have
\begin{align*}
 &\E\bigl[\ind{\tau_k\leq1}\ind{y_k=y}F_{k-1}(y_1,\dots,y_{k-1},\tau_k)\bigr]\\
 &\qquad\overset{\mathclap{\text{(I)}}}{=}\int_0^1R\,\Prb(K(t)=k-1)\,\Prb(y_k=y)\,
 \E\bigl[F_{k-1}(y_1,\dots,y_{k-1},t)\bigr]\,\dd t\\
 &\qquad=\lambda_y\int_0^1\Prb(K(t)=k-1)\,
 \E\bigl[F_{k-1}(y_1,\dots,y_{k-1},t)\bigr]\,\dd t\\
 &\qquad\overset{\mathclap{\text{(II)}}}{=}\lambda_y\int_0^1\E\bigl[\ind{K(t)=k-1}\,
 F_{K(t)}(y_1,\dots,y_{K(t)},t)\bigr]\,\dd t,
\end{align*}
where step (I) holds as $\tau_k$, $y_k$ and $(y_1,\dots,y_{k-1})$ are
independent, and step (II) holds as $K(t)$ is independent of the types.
We sum over $k\geq1$. As all terms are nonnegative, we may exchange the sum
with the expectations and the integral, and $\sum_{k\geq1}\ind{K(t)=k-1}=1$.
\end{proof}

\subsection{Blocked arrivals}\label{app:blocked}

For an edge $e=(u,v)$ of $G$ and $t\in[0,1]$, let $D_e(t)$ be the event that
some arrival among $y_1,\dots,y_{K(t)}$ has type $f^+$ or $f^-$ for an edge $f$
incident to $u$ or $v$. An arrival of type $e^+$ or $e^-$ that follows
$y_1,\dots,y_{K(t)}$ is blocked if and only if $D_e(t)$ occurs, so
$P_e(t)=1-\Prb(D_e(t))$.

\begin{lemma}\label{lem:blocked}
\begin{enumerate}[(a),nosep]
\item Let $e=(u,v)$ be an edge of $G$ and $t\in[0,1]$. The event $D_e(t)$ is
independent of the numbers $X_w(t)$, $w\in V$, and
\[
 P_e(t)=\exp\bigl(-2\lambda t(\deg(u)+\deg(v)-1)\bigr)\geq1-4\Delta_G\lambda t.
\]
\item The expected number of blocked arrivals among $y_1,\dots,y_{K(1)}$ is at
most $4\Delta_G\lambda^2m$.
\end{enumerate}
\end{lemma}

\begin{proof}
\emph{Part (a).} As $G$ is simple, exactly $\deg(u)+\deg(v)-1$ edges are
incident to $u$ or $v$, and each of them has two edge types of rate
$\lambda$. As the Poisson processes of the types are independent, the number
$L$ of arrivals with these types among $y_1,\dots,y_{K(t)}$ is independent of
the numbers $X_w(t)$, $w\in V$, and it has distribution
$\Pois(2\lambda t(\deg(u)+\deg(v)-1))$. As
$D_e(t)=\{L\geq1\}$, we obtain the independence, and by $e^{-x}\geq1-x$,
\begin{align*}
 P_e(t)=\Prb(L=0)
 &=\exp\bigl(-2\lambda t(\deg(u)+\deg(v)-1)\bigr)\\
 &\geq1-2\lambda t(\deg(u)+\deg(v)-1)\\
 &\geq1-2\lambda t(2\Delta_G-1)\geq1-4\Delta_G\lambda t.
\end{align*}

\emph{Part (b).} Only edge arrivals are blocked. We fix an edge type $e^s$,
and we let $F_\ell(y_1,\dots,y_\ell,t)$ be $1$ if some $y_i$ is $f^+$ or $f^-$
for an edge $f$ incident to an endpoint of $e$, and $0$ otherwise. If
$y_k=e^s$, the $k$-th arrival is blocked if and only if
$F_{k-1}(y_1,\dots,y_{k-1},\tau_k)=1$. Moreover,
$F_{K(t)}(y_1,\dots,y_{K(t)},t)$ is the indicator of $D_e(t)$. By
\cref{lem:poisson} and part~(a), the expected number of blocked arrivals of
type $e^s$ among $y_1,\dots,y_{K(1)}$ is
\[
 \lambda\int_0^1\bigl(1-P_e(t)\bigr)\,\dd t
 \leq\lambda\int_0^14\Delta_G\lambda t\,\dd t=2\Delta_G\lambda^2.
\]
We sum over the $2m$ edge types.
\end{proof}

\subsection{Proof of Lemma~\ref{lem:transfer} and Corollary~\ref{cor:transfer}}\label{app:transfer}

\transferlemma*

\begin{proof}
Both models take their types from $y_1,y_2,\dots$, and $K(1)$ is independent
of the types.

On every sequence of types,
$M_\pi(\ell+1)-M_\pi(\ell)\in\{0,1\}$ for every $\ell\geq0$, so
$|M_\pi(T)-M_\pi(K(1))|\leq|T-K(1)|$. The number $K(1)$ has mean and variance
$R$, and $R\leq T<R+1$. By the Cauchy--Schwarz inequality,
\begin{equation}\label{eq:poisson-count}
\begin{aligned}
 \bigl|\E[M_\pi(T)]-\E[M_\pi(K(1))]\bigr|
 &\leq\E|T-K(1)|\leq\E|K(1)-R|+(T-R)\\
 &\leq\sqrt{\Var(K(1))}+1=\sqrt R+1.
\end{aligned}
\end{equation}

We fix a sequence of types. In the adjusted model,
$\pi$ takes the same decisions as in $I(G,\lambda)$, except that it rejects
every blocked arrival. We show by induction on $\ell$ that its decisions on
the first $\ell$ arrivals are feasible in the adjusted model, and that every
vertex that $\pi$ uses there is also used in $I(G,\lambda)$. Suppose that
$\pi$ matches arrival $\ell+1$ in the adjusted model. Then this arrival is not
blocked, and $\pi$ matches it to a neighbor $w$ that is free in
$I(G,\lambda)$, so $w$ is also free in the adjusted model. Both models then
use $w$. Hence $\pi$ is a policy of the adjusted model, and
\[
 \widehat M_\pi(\ell)=M_\pi(\ell)-Z_\ell,
\]
where $Z_\ell$ is the number of blocked arrivals among the first $\ell$ that
$\pi$ matches in $I(G,\lambda)$. By \cref{lem:blocked}(b),
$0\leq\E[Z_{K(1)}]\leq4\Delta_G\lambda^2m$, so
\[
 \E[M_\pi(K(1))]-4\Delta_G\lambda^2m\leq\E[\widehat M_\pi]
 \leq\E[M_\pi(K(1))].
\]
We combine this with \eqref{eq:poisson-count}.
\end{proof}

\transfercor*

\begin{proof}
\emph{Lower bound.} Let $\pi$ be a deterministic policy of the adjusted model,
and let $\pi'$ take the decisions of $\pi$, except that it rejects every
blocked arrival. Blocked arrivals have no neighbors in the adjusted model, so
$\widehat M_{\pi'}=\widehat M_\pi$. On every sequence of types, $\pi'$ matches
the same arrivals to the same vertices in $I(G,\lambda)$ as in the adjusted
model, so $\pi'$ is a policy for $I(G,\lambda)$. By \cref{lem:transfer},
\begin{align*}
 \E[\widehat M_\pi]=\E[\widehat M_{\pi'}]
 &\leq\E[M_{\pi'}(T)]+\sqrt R+1\\
 &\leq\OPT(I(G,\lambda))+\sqrt R+1.
\end{align*}
We take the supremum over $\pi$.

\emph{Upper bound.} By \cref{sec:iid}, $\OPT(I(G,\lambda))=\E[M_\pi(T)]$ for
a deterministic policy $\pi$. By the proof of \cref{lem:transfer}, $\pi$ is
also a policy of the adjusted model, and
\begin{align*}
 \OPT(I(G,\lambda))=\E[M_\pi(T)]
 &\overset{\mathclap{\text{(I)}}}{\leq}\E[\widehat M_\pi]+4\Delta_G\lambda^2m+\sqrt R+1\\
 &\leq\widehat\OPT+4\Delta_G\lambda^2m+\sqrt R+1,
\end{align*}
where step (I) is by \cref{lem:transfer}.
\end{proof}

\subsection{Proof of Lemma~\ref{lem:accounting}}\label{app:accounting}

\accountinglemma*

\begin{proof}
For $d\in\{0,1,2\}$ and $t\in[0,1]$, let $f_d(t)=\E[\min\{\xi,d\}]$, where
$\xi$ has distribution $\Pois(1-t)$. As
$\min\{\xi,d\}-\min\{\xi,d-1\}=\ind{\xi\geq d}$,
\begin{equation}\label{eq:fh}
\begin{aligned}
 h_d(t)=\Prb(\xi\leq d-1)
 &=1-\E\bigl[\min\{\xi,d\}-\min\{\xi,d-1\}\bigr]\\
 &=1+f_{d-1}(t)-f_d(t)\qquad\text{for }d\in\{1,2\}.
\end{aligned}
\end{equation}
We have $f_0=0$, $f_1(t)=1-e^{t-1}$ and $f_2(t)=2-(3-t)e^{t-1}$, so for
$d\in\{0,1,2\}$,
\[
 \frac{\dd}{\dd t}f_d(t)=-h_d(t),\qquad f_d(1)=0,\qquad\text{and}\qquad
 f_2(0)=2-\frac3e.
\]

We fix the types and the times, with $K(1)<\infty$. For
$\ell\geq0$ and $u\in V$, let $S_{u,\ell}$ be the set of free vertices of the
pair of $u$ after $\pi$ processes $y_1,\dots,y_\ell$ in the adjusted model, and
let $\Psi_\ell(t)=\sum_{u\in V}f_{|S_{u,\ell}|}(t)$. With $\tau_0=0$, the
intervals $[\tau_k,\min\{\tau_{k+1},1\}]$, $0\leq k\leq K(1)$, cover $[0,1]$,
and $S_{u,K(t)}=S_{u,k}$ for $t\in[\tau_k,\tau_{k+1})$. As
$\frac{\dd}{\dd t}f_d=-h_d$,
\begin{equation}\label{eq:telescoping}
\begin{aligned}
 \Psi_{K(1)}(1)-\Psi_0(0)
 &=\sum_{k=0}^{K(1)}\Bigl[\Psi_k\bigl(\min\{\tau_{k+1},1\}\bigr)
 -\Psi_k(\tau_k)\Bigr]
 +\sum_{k=1}^{K(1)}\Bigl[\Psi_k(\tau_k)-\Psi_{k-1}(\tau_k)\Bigr]\\
 &=-\sum_{k=0}^{K(1)}\int_{\tau_k}^{\min\{\tau_{k+1},1\}}
 \sum_{u\in V}h_{|S_{u,k}|}(t)\,\dd t
 +\sum_{k=1}^{K(1)}\Bigl[\Psi_k(\tau_k)-\Psi_{k-1}(\tau_k)\Bigr]\\
 &=-\int_0^1\sum_{u\in V}h_{|S_{u,K(t)}|}(t)\,\dd t
 +\sum_{k=1}^{K(1)}\Bigl[\Psi_k(\tau_k)-\Psi_{k-1}(\tau_k)\Bigr].
\end{aligned}
\end{equation}
Here $\Psi_{K(1)}(1)=0$ and $\Psi_0(0)=nf_2(0)=(2-3/e)n$. For
$1\leq k\leq K(1)$, let
$\rho_k=\widehat M_\pi(k)-\widehat M_\pi(k-1)+\Psi_k(\tau_k)-\Psi_{k-1}(\tau_k)$.
We sum $\rho_k$ over $k$. As $\widehat M_\pi(0)=0$, \eqref{eq:telescoping}
gives
\begin{equation}\label{eq:potential}
\begin{aligned}
 \widehat M_\pi
 &=\sum_{k=1}^{K(1)}\rho_k
 -\sum_{k=1}^{K(1)}\Bigl[\Psi_k(\tau_k)-\Psi_{k-1}(\tau_k)\Bigr]\\
 &=\sum_{k=1}^{K(1)}\rho_k+\Psi_0(0)-\Psi_{K(1)}(1)
 -\int_0^1\sum_{u\in V}h_{|S_{u,K(t)}|}(t)\,\dd t\\
 &=\Bigl(2-\frac3e\Bigr)n+\sum_{k=1}^{K(1)}\rho_k
 -\int_0^1\sum_{u\in V}h_{|S_{u,K(t)}|}(t)\,\dd t.
\end{aligned}
\end{equation}

If $\pi$ rejects the $k$-th arrival, then $\rho_k=0$.
Suppose that $\pi$ matches it to a vertex of the pair of $w\in V$, and let
$d=|S_{w,k-1}|$. Only the set of free vertices of this pair changes, and it
loses one vertex, so $\rho_k=1+f_{d-1}(\tau_k)-f_d(\tau_k)=h_d(\tau_k)$ by
\eqref{eq:fh}. Vertex arrivals are never blocked, and $\pi$ is greedy, so
$\pi$ rejects an arrival of the vertex type $u$ only if $S_{u,k-1}=\emptyset$.
As $h_0=0$, we have $\rho_k=h_{|S_{u,k-1}|}(\tau_k)$ whenever $y_k=u$. The
events $\{k\leq K(1)\}$ and $\{\tau_k\leq1\}$ coincide, so \eqref{eq:potential}
becomes
\begin{align*}
 \widehat M_\pi=\Bigl(2-\frac3e\Bigr)n
 &+\sum_{u\in V}\Bigl[\sum_{k\geq1:\,\tau_k\leq1}\ind{y_k=u}\,
 h_{|S_{u,k-1}|}(\tau_k)-\int_0^1h_{|S_{u,K(t)}|}(t)\,\dd t\Bigr]\\
 &+\sum_{e\in E}\sum_{s\in\{+,-\}}\sum_{k\geq1:\,\tau_k\leq1}
 \ind{y_k=e^s}\,\rho_k.
\end{align*}

We take expectations. All terms have finite expectations, as
$0\leq\rho_k\leq1$ and $\E[K(1)]=R$. We apply \cref{lem:poisson} twice.
\begin{itemize}[leftmargin=*,itemsep=1pt]
\item We take $y=u$, of rate $\lambda_u=1$, and
$F_\ell(y_1,\dots,y_\ell,t)=h_{|S_{u,\ell}|}(t)$. Exchanging the expectation and
the integral,
\[
 \E\Bigl[\sum_{k\geq1:\,\tau_k\leq1}\ind{y_k=u}\,h_{|S_{u,k-1}|}(\tau_k)\Bigr]
 =\E\Bigl[\int_0^1h_{|S_{u,K(t)}|}(t)\,\dd t\Bigr].
\]
\item We take $y=e^s$, of rate $\lambda$, and we let
$F_\ell(y_1,\dots,y_\ell,t)$ be $h_d(t)$ if $\pi$ matches an arrival of type
$e^s$ that follows $(y_1,\dots,y_\ell)$ to a vertex of a pair with $d$ free
vertices, and $0$ if $\pi$ rejects this arrival. Then
$F_{k-1}(y_1,\dots,y_{k-1},\tau_k)=\rho_k$ if $y_k=e^s$, and
$\E[F_{K(t)}(y_1,\dots,y_{K(t)},t)]=\sum_{d=1}^2h_d(t)\,P_e(t)\,Q^s_{e,d}(t)$
by the definitions of $P_e(t)$ and $Q^s_{e,d}(t)$. So
\[
 \E\Bigl[\sum_{k\geq1:\,\tau_k\leq1}\ind{y_k=e^s}\,\rho_k\Bigr]
 =\lambda\sum_{d=1}^2\int_0^1h_d(t)\,P_e(t)\,Q^s_{e,d}(t)\,\dd t.
\]
\end{itemize}
Since $h_0=0$, we obtain the lemma.
\end{proof}

\subsection{Net gains of an edge}\label{app:gains}

Let $\pi$ be a deterministic policy of the adjusted model, $t\in[0,1]$ and
$u\in V$. If the pair of $u$ has exactly one free vertex after $\pi$ processes
$y_1,\dots,y_{K(t)}$ in the adjusted model, we let $\theta_t(u)$ be the sign
of this vertex; otherwise, we let $\theta_t(u)=+$. For an edge $e=(u,v)$ of
$G$, let
\[
 \Gamma_e(t)=2h_2(t)\ind{\min\{X_u(t),X_v(t)\}=0}
 +h_1(t)\ind{\min\{X_u(t),X_v(t)\}=1}.
\]
In the adjusted model, we let $\pi_\sigma$ reject every blocked arrival and
follow \cref{alg:cut-policy} on the other arrivals, with the free vertices of
the adjusted model. Then $\pi_\sigma$ is a greedy policy of the adjusted
model.

\begin{lemma}\label{lem:gains}
Let $t\in[0,1]$, and let $e=(u,v)$ be an edge of $G$.
\begin{enumerate}[(a),nosep]
\item For every deterministic greedy policy $\pi$ of the adjusted model,
\[
 P_e(t)\sum_{s\in\{+,-\}}\sum_{d=1}^2h_d(t)\,Q^s_{e,d}(t)
 \leq\E[\Gamma_e(t)]
 +h_1(t)\,\Prb\bigl(X_u(t)=X_v(t)=1,\ \theta_t(u)\neq\theta_t(v)\bigr).
\]
\item For the policy $\pi_\sigma$ of a cut $\sigma$ of $G$,
\[
 \sum_{s\in\{+,-\}}\sum_{d=1}^2h_d(t)\,Q^s_{e,d}(t)
 =\E[\Gamma_e(t)]+h_1(t)\,t^2e^{-2t}\,\ind{\sigma(u)\neq\sigma(v)}.
\]
\item We have
\[
 \int_0^1\E[\Gamma_e(t)]\,\dd t=\frac1e+\frac9{e^2}
 \qquad\text{and}\qquad
 \int_0^1h_1(t)\,t^2e^{-2t}\,\dd t=\frac2e-\frac5{e^2}.
\]
\end{enumerate}
\end{lemma}

\begin{proof}
For a sign $s$, we consider an arrival of type $e^s$ that follows
$y_1,\dots,y_{K(t)}$. Let $L_s=h_d(t)$ if this arrival is not blocked and
$\pi$ matches it to a vertex of a pair with $d$ free vertices, and let $L_s=0$
otherwise. By the definitions of $P_e(t)$ and $Q^s_{e,d}(t)$,
\[
 \E[L_s]=P_e(t)\sum_{d=1}^2h_d(t)\,Q^s_{e,d}(t).
\]
On $D_e(t)$, the arrival is blocked, and $L_+=L_-=0$. We show that if $D_e(t)$
does not occur, then
\begin{equation}\label{eq:edge-pointwise}
 L_++L_-\leq\Gamma_e(t)
 +h_1(t)\ind{X_u(t)=X_v(t)=1,\ \theta_t(u)\neq\theta_t(v)},
\end{equation}
with equality for $\pi=\pi_\sigma$.

Suppose that $D_e(t)$ does not occur. Then
the arrivals among $y_1,\dots,y_{K(t)}$ with a neighbor in the pair of $u$ are
the $X_u(t)$ vertex arrivals of $u$, which are never blocked. As $\pi$ is
greedy, the pair of $u$ has $\max\{2-X_u(t),0\}$ free vertices after them, and
if $X_u(t)=1$, its free vertex is $u^{\theta_t(u)}$. The policy $\pi_\sigma$
matches the first vertex arrival of $u$ to $u^{-\sigma(u)}$, so
$\theta_t(u)=\sigma(u)$ if $X_u(t)=1$. The same holds for $v$.

The arrival of type $e^s$ is then not blocked. Let $\gamma_s$
be the largest $h_d(t)$ over the free vertices $w^s$ with $w\in\{u,v\}$, where
$d$ is the number of free vertices of the pair of $w$, and let $\gamma_s=0$ if
neither $u^s$ nor $v^s$ is free. If $\pi$ matches the arrival, it uses one of
these vertices, so $L_s\leq\gamma_s$. We have $h_2\geq h_1$, as
$h_d(t)=\Prb(\Pois(1-t)\leq d-1)$. The policy $\pi_\sigma$ matches the arrival
whenever $u^s$ or $v^s$ is free, and it prefers a vertex whose pair has two
free vertices, so $L_s=\gamma_s$ for $\pi=\pi_\sigma$. The free vertices of
the pairs of $u$ and $v$ give
\begin{align*}
 \gamma_++\gamma_-&=\begin{cases}
 2h_2(t)&\text{if }\min\{X_u(t),X_v(t)\}=0,\\
 h_1(t)\bigl(1+\ind{X_u(t)=X_v(t)=1,\ \theta_t(u)\neq\theta_t(v)}\bigr)
 &\text{if }\min\{X_u(t),X_v(t)\}=1,\\
 0&\text{if }\min\{X_u(t),X_v(t)\}\geq2
 \end{cases}\\
 &=\Gamma_e(t)+h_1(t)\ind{X_u(t)=X_v(t)=1,\ \theta_t(u)\neq\theta_t(v)}.
\end{align*}
As $L_+\leq\gamma_+$ and $L_-\leq\gamma_-$, with equality for $\pi=\pi_\sigma$,
this proves \eqref{eq:edge-pointwise}.

\emph{Part (a).} As $\Gamma_e(t)\geq0$ and $L_+=L_-=0$ on $D_e(t)$,
\eqref{eq:edge-pointwise} also holds on $D_e(t)$. We take expectations.

\emph{Part (b).} For $\pi=\pi_\sigma$, if $D_e(t)$ does not occur and
$X_u(t)=X_v(t)=1$, then $\theta_t(u)=\sigma(u)$ and $\theta_t(v)=\sigma(v)$.
So
\[
 L_++L_-=\Bigl(\Gamma_e(t)+h_1(t)\ind{X_u(t)=X_v(t)=1}\ind{\sigma(u)\neq\sigma(v)}
 \Bigr)\bigl(1-\ind{D_e(t)}\bigr).
\]
The first factor is a function of $X_u(t)$ and $X_v(t)$, which are
independent of $D_e(t)$ by \cref{lem:blocked}(a). As $X_u(t)$ and $X_v(t)$
are independent with distribution $\Pois(t)$,
$\Prb(X_u(t)=X_v(t)=1)=t^2e^{-2t}$. We take expectations and divide by
$P_e(t)>0$.

\emph{Part (c).} As $X_u(t)$ and $X_v(t)$ are independent with distribution
$\Pois(t)$,
\begin{align*}
 \Prb\bigl(\min\{X_u(t),X_v(t)\}=0\bigr)
 &=1-\bigl(1-e^{-t}\bigr)^2=2e^{-t}-e^{-2t},\\
 \Prb\bigl(\min\{X_u(t),X_v(t)\}=1\bigr)
 &=\bigl(1-e^{-t}\bigr)^2-\bigl(1-(1+t)e^{-t}\bigr)^2
 =2te^{-t}-t(2+t)e^{-2t}.
\end{align*}
With $h_1(t)=e^{t-1}$, $h_2(t)=(2-t)e^{t-1}$ and
$\int_0^1t^2e^{-t}\,\dd t=2-5/e$,
\begin{align*}
 \int_0^1\E[\Gamma_e(t)]\,\dd t
 &=\int_0^1\Bigl(2h_2(t)\bigl(2e^{-t}-e^{-2t}\bigr)
 +h_1(t)\bigl(2te^{-t}-t(2+t)e^{-2t}\bigr)\Bigr)\,\dd t\\
 &=\frac1e\int_0^1\bigl(8-2t-(4+t^2)e^{-t}\bigr)\,\dd t\\
 &=\frac1e\Bigl(7-4\Bigl(1-\frac1e\Bigr)-\Bigl(2-\frac5e\Bigr)\Bigr)
 =\frac1e+\frac9{e^2},\\
 \int_0^1h_1(t)\,t^2e^{-2t}\,\dd t
 &=\frac1e\int_0^1t^2e^{-t}\,\dd t=\frac2e-\frac5{e^2}.\qedhere
\end{align*}
\end{proof}

\subsection{Proof of Lemma~\ref{lem:cutvalue}}\label{app:cutvalue}

\cutvaluelemma*

\begin{proof}
As $\pi_\sigma$ is a greedy policy of the adjusted model and $h_0=0$, by
\cref{lem:accounting},
\[
 \E[\widehat M_{\pi_\sigma}]
 =\Bigl(2-\frac3e\Bigr)n+\lambda\sum_{e\in E}\int_0^1P_e(t)
 \sum_{s\in\{+,-\}}\sum_{d=1}^2h_d(t)\,Q^s_{e,d}(t)\,\dd t.
\]
By \cref{lem:gains}(b) and (c), every edge $e=(u,v)$ of $G$ satisfies
\[
 \int_0^1\sum_{s\in\{+,-\}}\sum_{d=1}^2h_d(t)\,Q^s_{e,d}(t)\,\dd t
 =\frac1e+\frac9{e^2}+\Bigl(\frac2e-\frac5{e^2}\Bigr)\ind{\sigma(u)\neq\sigma(v)}.
\]
We sum over the edges and compare with the definition of $\Phi_G$. As
$0\leq h_d(t)\leq1$, $Q^s_{e,1}(t)+Q^s_{e,2}(t)\leq1$ and
$0\leq1-P_e(t)\leq4\Delta_G\lambda t$ by \cref{lem:blocked}(a),
\begin{align*}
 0\leq\Phi_G(|\delta_G(\sigma)|)-\E[\widehat M_{\pi_\sigma}]
 &=\lambda\sum_{e\in E}\int_0^1\bigl(1-P_e(t)\bigr)
 \sum_{s\in\{+,-\}}\sum_{d=1}^2h_d(t)\,Q^s_{e,d}(t)\,\dd t\\
 &\leq\lambda m\int_0^18\Delta_G\lambda t\,\dd t=4\Delta_G\lambda^2m.\qedhere
\end{align*}
\end{proof}

\subsection{Proof of Lemma~\ref{lem:blowup}}\label{app:blowups}

\blowuplemma*

\begin{proof}
For $u\in V$, let $a_u$ be the number of copies of $u$ in $A$. It has distribution
$\Bin(B,p)$.

\emph{Lower bound.} Let $\theta$ be a maximum cut of $\widetilde G$. Its
restriction to $A$ is a cut of $\widetilde G[A]$, and it cuts every edge of
$\delta_{\widetilde G}(\theta)$ with both endpoints in $A$. Every edge has both
endpoints in $A$ with probability $p^2$, so
$\E[C(\widetilde G[A])]\geq p^2|\delta_{\widetilde G}(\theta)|
=p^2C(\widetilde G)$.

\emph{Upper bound.} We fix $A$ and a maximum cut $\theta$ of
$\widetilde G[A]$. As in the proof of $C(\widetilde G)=B^2C(G)$ in
\cref{sec:adaptivity}, we define a random cut $\sigma'$ of $G$. Every
$u\in V$ with $a_u\geq1$ receives the sign under $\theta$ of a uniformly
random copy of $u$ in $A$, independently for different $u$, and every other
$u$ receives the sign $+$. For an edge $(u,v)\in E$, the graph
$\widetilde G[A]$ has $a_ua_v$ edges between the copies of $u$ and of $v$, and
each of them joins the two chosen copies with probability $1/(a_ua_v)$. Taking
the expectation over $\sigma'$,
\begin{align*}
 C(\widetilde G[A])=\bigl|\delta_{\widetilde G[A]}(\theta)\bigr|
 &=\E\Bigl[\sum_{(u,v)\in E}a_ua_v\ind{\sigma'(u)\neq\sigma'(v)}\Bigr]\\
 &\leq\max_{\sigma'}\sum_{(u,v)\in E}a_ua_v\ind{\sigma'(u)\neq\sigma'(v)},
\end{align*}
where the maximum is over the cuts $\sigma'$ of $G$. For every cut $\sigma'$
of $G$, we have $B^2|\delta_G(\sigma')|\leq B^2C(G)=C(\widetilde G)$, so
\begin{align*}
 \sum_{(u,v)\in E}a_ua_v\ind{\sigma'(u)\neq\sigma'(v)}
 &\leq p^2B^2|\delta_G(\sigma')|+\sum_{(u,v)\in E}\bigl|a_ua_v-p^2B^2\bigr|\\
 &\leq p^2C(\widetilde G)+\sum_{(u,v)\in E}\bigl|a_ua_v-p^2B^2\bigr|.
\end{align*}
For an edge $(u,v)\in E$, as $0\leq a_u\leq B$ and
$\Var(a_u)=Bp(1-p)\leq B/4$, by the Cauchy--Schwarz inequality,
\begin{align*}
 \E\bigl|a_ua_v-p^2B^2\bigr|
 &=\E\bigl|a_u(a_v-pB)+pB(a_u-pB)\bigr|\\
 &\leq B\,\E|a_v-pB|+B\,\E|a_u-pB|\\
 &\leq B\sqrt{\Var(a_v)}+B\sqrt{\Var(a_u)}\leq B^{3/2}.
\end{align*}
We take expectations and sum over the $m$ edges of $G$.
\end{proof}

\subsection{Proof of Lemma~\ref{lem:gap}}\label{app:gap}

\gaplemma*

\begin{proof}
As $\pi_\sigma$ is a policy of the adjusted model,
$\E[\widehat M_{\pi_\sigma}]\leq\widehat\OPT$. We apply the lemmas of this
section to $\widetilde G$.

Let $\pi$ be a deterministic policy of the adjusted
model. By \cref{prop:greedy}, where a blocked arrival counts as an arrival of a
type without neighbors, there is a deterministic greedy policy $\pi'$ of the
adjusted model with $\E[\widehat M_\pi]\leq\E[\widehat M_{\pi'}]$. Let
$t\in[0,1]$, let $\theta_t$ be the signs of $\pi'$, and let $A(t)$ be the set
of vertices $w$ of $\widetilde G$ with $X_w(t)=1$. The restriction of
$\theta_t$ to $A(t)$ is a cut of $\widetilde G[A(t)]$. As the numbers $X_w(t)$
are independent with distribution $\Pois(t)$, the set $A(t)$ contains every
vertex of $\widetilde G$ independently with probability $te^{-t}$. We have
\begin{align*}
 &\qquad\qquad\sum_{e\in E(\widetilde G)}P_e(t)\sum_{s\in\{+,-\}}\sum_{d=1}^2
 h_d(t)\,Q^s_{e,d}(t)\\
 &\qquad\overset{\mathclap{\text{(I)}}}{\leq}\sum_{e\in E(\widetilde G)}\E[\Gamma_e(t)]
 +h_1(t)\sum_{(u,v)\in E(\widetilde G)}
 \Prb\bigl(X_u(t)=X_v(t)=1,\ \theta_t(u)\neq\theta_t(v)\bigr)\\
 &\qquad\overset{\mathclap{\text{(II)}}}{=}\sum_{e\in E(\widetilde G)}\E[\Gamma_e(t)]
 +h_1(t)\,\E\bigl[|\delta_{\widetilde G[A(t)]}(\theta_t)|\bigr]\\
 &\qquad\leq\sum_{e\in E(\widetilde G)}\E[\Gamma_e(t)]
 +h_1(t)\,\E\bigl[C(\widetilde G[A(t)])\bigr]\\
 &\qquad\overset{\mathclap{\text{(III)}}}{\leq}\sum_{e\in E(\widetilde G)}\E[\Gamma_e(t)]
 +h_1(t)\bigl(t^2e^{-2t}C(\widetilde G)+mB^{3/2}\bigr),
\end{align*}
where step (I) is by \cref{lem:gains}(a) for $\pi'$, summed over the edges of
$\widetilde G$, step (II) holds as the edges of $\widetilde G[A(t)]$ are the
edges $(u,v)$ of $\widetilde G$ with $X_u(t)=X_v(t)=1$, and step (III) is by
\cref{lem:blowup} with $p=te^{-t}$.

We integrate over $t$. By \cref{lem:accounting} for $\pi'$,
\begin{align*}
 \E[\widehat M_{\pi'}]
 &=\Bigl(2-\frac3e\Bigr)nB+\lambda\int_0^1\sum_{e\in E(\widetilde G)}P_e(t)
 \sum_{s\in\{+,-\}}\sum_{d=1}^2h_d(t)\,Q^s_{e,d}(t)\,\dd t\\
 &\overset{\mathclap{\text{(I)}}}{\leq}\Bigl(2-\frac3e\Bigr)nB
 +\Bigl(\frac1e+\frac9{e^2}\Bigr)\lambda mB^2
 +\Bigl(\frac2e-\frac5{e^2}\Bigr)\lambda C(\widetilde G)+\lambda mB^{3/2}\\
 &=\Phi_{\widetilde G}(C(\widetilde G))+\lambda mB^{3/2},
\end{align*}
where step (I) is by \cref{lem:gains}(c) and
$\int_0^1h_1(t)\,\dd t=1-e^{-1}\leq1$.

As $\pi$ is arbitrary,
$\widehat\OPT\leq\Phi_{\widetilde G}(C(\widetilde G))+\lambda mB^{3/2}$. By
\cref{lem:cutvalue} for $\widetilde G$ and $\sigma$, with
$|\delta_{\widetilde G}(\sigma)|=C(\widetilde G)$,
\begin{align*}
 \E[\widehat M_{\pi_\sigma}]
 &\geq\Phi_{\widetilde G}(C(\widetilde G))-4(B\Delta_G)\lambda^2(mB^2)\\
 &\geq\widehat\OPT-\lambda mB^{3/2}-4\Delta_G\lambda^2mB^3.\qedhere
\end{align*}
\end{proof}

\subsection{Proof of Theorem~\ref{thm:main}}\label{app:main}

\mainthm*

\begin{proof}
Let $G$ be cubic, so that $\Delta_G=3$ and $m=3n/2$, and let $\sigma$ be a
maximum cut of $\widetilde G$. The instance
$I(\widetilde G,\lambda)$ has total rate $R=nB+2\lambda mB^2$, and
$\Delta_{\widetilde G}=3B$. By the upper bounds in \cref{cor:transfer} for
$\widetilde G$, \cref{lem:gap} and \cref{lem:cutvalue} for $\widetilde G$, we
have
\begin{align*}
 \OPT(I(\widetilde G,\lambda))
 &\leq\widehat\OPT+12\lambda^2mB^3+\sqrt R+1\\
 &\leq\E[\widehat M_{\pi_\sigma}]+\lambda mB^{3/2}+24\lambda^2mB^3+\sqrt R+1\\
 &\leq\Phi_{\widetilde G}(C(\widetilde G))+\lambda mB^{3/2}
 +24\lambda^2mB^3+\sqrt R+1,
\end{align*}
and by the lower bounds in these results, we have
\begin{align*}
 \OPT(I(\widetilde G,\lambda))
 &\geq\widehat\OPT-\sqrt R-1\\
 &\geq\E[\widehat M_{\pi_\sigma}]-\sqrt R-1\\
 &\geq\Phi_{\widetilde G}(C(\widetilde G))-12\lambda^2mB^3-\sqrt R-1.
\end{align*}
As $C(\widetilde G)=B^2C(G)$,
\begin{equation}\label{eq:estimate}
 \bigl|\OPT(I(\widetilde G,\lambda))-\Phi_{\widetilde G}\bigl(B^2C(G)\bigr)\bigr|
 \leq\lambda mB^2\Bigl(24\lambda B+\frac1{\sqrt B}\Bigr)+\sqrt R+1.
\end{equation}

We fix a rational number $c>0$, an integer $B\geq1$ and $\varepsilon_0>0$ with
$24c+1/\sqrt B+2\varepsilon_0/c\leq\eta_0/80$, and an integer $N_0$ such that
$(\sqrt{2mB}+2)/(cmB)\leq\eta_0/80$ for all $m\geq N_0$.

Let $\mathrm{ALG}$ be an algorithm that approximates the online optimum within
a factor $1\pm\varepsilon_0$ on the instances of \cref{thm:main}. We use
$\mathrm{ALG}$ as an oracle to approximate $C(G)$ within a factor
$1\pm\eta_0$ for a simple cubic graph $G$. If $m<N_0$, we compute $C(G)$ by
enumerating the cuts of $G$. Otherwise, we let $\lambda=c/B$, so that
$R=(1+3c)nB$. Then $I(\widetilde G,\lambda)$ is an instance of
\cref{thm:main} with $2nB$ offline vertices. We query $\mathrm{ALG}$ on it and
obtain $\bar V$. We output a rational number $\bar C$ with
$|\Phi_{\widetilde G}(B^2\bar C)-\bar V|\leq1$, which we compute from a
rational approximation of $e$. As $c$ and $B$ are constants, the algorithm
runs in polynomial time.

It remains to show that $|\bar C-C(G)|\leq\eta_0C(G)$ if $m\geq N_0$. Let
$V^*=\OPT(I(\widetilde G,\lambda))$. As $V^*\leq2nB\leq2mB$, the guarantee of
$\mathrm{ALG}$ gives $|\bar V-V^*|\leq2\varepsilon_0mB$. So
\begin{equation}\label{eq:error}
\begin{aligned}
 \bigl|\bar V-\Phi_{\widetilde G}\bigl(B^2C(G)\bigr)\bigr|+1
 &\leq|\bar V-V^*|+\bigl|V^*-\Phi_{\widetilde G}\bigl(B^2C(G)\bigr)\bigr|+1\\
 &\overset{\mathclap{\text{(I)}}}{\leq}2\varepsilon_0mB+cmB\Bigl(24c+\frac1{\sqrt B}\Bigr)
 +\sqrt{2mB}+2\\
 &=cmB\Bigl(24c+\frac1{\sqrt B}+\frac{2\varepsilon_0}c
 +\frac{\sqrt{2mB}+2}{cmB}\Bigr)\\
 &\overset{\mathclap{\text{(II)}}}{\leq}cmB\Bigl(\frac{\eta_0}{80}+\frac{\eta_0}{80}\Bigr)
 =\frac{c\eta_0mB}{40},
\end{aligned}
\end{equation}
where step (I) is by \eqref{eq:estimate} with $\lambda=c/B$ and $R\leq2mB$, and
step (II) is by the choice of the constants. As $2/e-5/e^2>1/20$,
\begin{align*}
 \frac{cB}{20}\,\bigl|\bar C-C(G)\bigr|
 &\leq\Bigl(\frac2e-\frac5{e^2}\Bigr)cB\,\bigl|\bar C-C(G)\bigr|\\
 &\overset{\mathclap{\text{(I)}}}{=}\bigl|\Phi_{\widetilde G}(B^2\bar C)
 -\Phi_{\widetilde G}\bigl(B^2C(G)\bigr)\bigr|\\
 &\leq\bigl|\Phi_{\widetilde G}(B^2\bar C)-\bar V\bigr|
 +\bigl|\bar V-\Phi_{\widetilde G}\bigl(B^2C(G)\bigr)\bigr|\\
 &\overset{\mathclap{\text{(II)}}}{\leq}\bigl|\bar V-\Phi_{\widetilde G}\bigl(B^2C(G)\bigr)\bigr|+1
 \leq\frac{c\eta_0mB}{40},
\end{align*}
where step (I) is by the definition of $\Phi_{\widetilde G}$ and $\lambda=c/B$,
and step (II) holds as $|\Phi_{\widetilde G}(B^2\bar C)-\bar V|\leq1$.
A uniformly random cut of $G$ cuts every edge with probability $1/2$, so
$C(G)\geq m/2$, and $|\bar C-C(G)|\leq\eta_0m/2\leq\eta_0C(G)$. Hence
approximating the online optimum within a factor $1\pm\varepsilon_0$ on the
instances of \cref{thm:main} is $\NP$-hard.
\end{proof}

\subsection{Proof of Theorem~\ref{thm:randomized}}\label{app:randomized}

\randthm*

\begin{proof}
We use the notation of the proof of \cref{thm:main}. The algorithm in that
proof queries $\mathrm{ALG}$ at most once, on $I(\widetilde G,\lambda)$, and its output
approximates $C(G)$ within a factor $1\pm\eta_0$ whenever the output $\bar V$
of the query satisfies $|\bar V-V^*|\leq2\varepsilon_0mB$.

Suppose that a randomized polynomial-time algorithm approximates the online
optimum within a factor $1\pm\varepsilon_0$ with probability at least $2/3$ on
the instances of \cref{thm:main}. We use it in place of $\mathrm{ALG}$. Then
$|\bar V-V^*|\leq\varepsilon_0V^*\leq2\varepsilon_0mB$ with probability at
least $2/3$, so we approximate $C(G)$ within a factor $1\pm\eta_0$ with
probability at least $2/3$. With the median of polynomially many independent
runs, the failure probability becomes smaller than any inverse polynomial.
Hence $\NP\subseteq\BPP$ by \cref{thm:maxcut}.

Suppose that a randomized polynomial-time algorithm $\mathcal A$ outputs, on
every instance $I$ of \cref{thm:main}, a polynomial-time online policy with
expected value at least $(1-\varepsilon_0/2)\OPT(I)$. It suffices to compute
in randomized polynomial time a number $\bar V$ such that
$|\bar V-V^*|\leq2\varepsilon_0mB$ with probability at least $2/3$. Let
$J_0=\lceil6/\varepsilon_0^2\rceil$. For $j=1,\dots,J_0$, we run $\mathcal A$
on $I(\widetilde G,\lambda)$ with fresh randomness, and we run the output
policy on simulated arrivals with $T$ independent types. Let $\zeta_j$ be the number of matches, and let
$\bar V=\frac1{J_0}\sum_{j=1}^{J_0}\zeta_j$.

Suppose first that every draw of the simulation is exact. Then
$\zeta_1,\dots,\zeta_{J_0}$ are independent and identically distributed in
$[0,2nB]$, so $\Var(\zeta_j)\leq(nB)^2$, and their mean $\mu$ satisfies
$(1-\varepsilon_0/2)V^*\leq\mu\leq V^*$. By Chebyshev's inequality,
\begin{align*}
 \Prb\bigl(|\bar V-\mu|\geq\varepsilon_0nB\bigr)
 &\leq\frac{\Var(\bar V)}{(\varepsilon_0nB)^2}
 =\frac{\Var(\zeta_1)}{J_0(\varepsilon_0nB)^2}\\
 &\leq\frac{(nB)^2}{J_0(\varepsilon_0nB)^2}\leq\frac16.
\end{align*}
On the complementary event, as $V^*\leq2nB$ and $n\leq m$,
\begin{align*}
 |\bar V-V^*|
 &\leq|\bar V-\mu|+V^*-\mu\\
 &<\varepsilon_0nB+\frac{\varepsilon_0V^*}2\leq2\varepsilon_0mB.
\end{align*}
We draw every type from fair random bits by rejection sampling, with
$\lceil\log_2(6J_0T)\rceil$ attempts per draw. Every attempt succeeds with
probability at least $1/2$. So the simulation coincides with an exact one
unless one of its $J_0T$ draws fails all its attempts, which has probability
at most $J_0T\cdot2^{-\lceil\log_2(6J_0T)\rceil}\leq1/6$. Hence
$|\bar V-V^*|\leq2\varepsilon_0mB$ with probability at least $2/3$. As $J_0$ is
a constant and the output policies run in polynomial time, we compute $\bar V$
in polynomial time.
\end{proof}

\section{Proofs for Section~\ref{sec:secretary}}\label{app:secretary}

\subsection{Proof of Lemma~\ref{lem:counting}}\label{app:counting}

\countinglemma*

\begin{proof}
For a real number $y$, let $y_+=\max\{y,0\}$. By the choice of
$\varepsilon$, and as $b$ is a positive multiple of $\varepsilon$, we have
$2A^2<\varepsilon\leq b\leq A$, so $A<1/2$. Hence the
three points $b-\varepsilon/4$, $b$ and $b+\varepsilon/4$ lie in $(0,1)$, and
\eqref{eq:value} holds at them.

For every real number $r$,
\begin{equation}\label{eq:second-diff}
\begin{aligned}
 \Bigl(b+\frac\varepsilon4-r\Bigr)_+-2(b-r)_+
 +\Bigl(b-\frac\varepsilon4-r\Bigr)_+
 &=\begin{cases}
 0&\text{if }r\geq b+\varepsilon/4,\\
 \varepsilon/4-(r-b)&\text{if }b\leq r<b+\varepsilon/4,\\
 \varepsilon/4-(b-r)&\text{if }b-\varepsilon/4\leq r<b,\\
 0&\text{if }r<b-\varepsilon/4
 \end{cases}\\
 &=\Bigl(\frac\varepsilon4-|b-r|\Bigr)_+.
\end{aligned}
\end{equation}

By the bounds on $r_S$ in
\cref{sec:secretary}, every solution $S$ satisfies $0\leq b-r_S\leq A^2/2$. As
$b$ and every $\sum_{i\in S}a_i$ are multiples of $\varepsilon$, the same
bounds give $|b-r_S|\geq\varepsilon/4$ for every other set $S\subseteq[n]$. We
have
\begin{align*}
 \Delta_1-\Delta_2
 &=\OPT\bigl(I(b+\tfrac\varepsilon4)\bigr)-2\OPT\bigl(I(b)\bigr)
 +\OPT\bigl(I(b-\tfrac\varepsilon4)\bigr)\\
 &\overset{\mathclap{\text{(I)}}}{=}\sum_{S\subseteq[n]}P_S\Bigl[\Bigl(b+\frac\varepsilon4-r_S\Bigr)_+
 -2(b-r_S)_++\Bigl(b-\frac\varepsilon4-r_S\Bigr)_+\Bigr]\\
 &\overset{\mathclap{\text{(II)}}}{=}\sum_{S\subseteq[n]}P_S\Bigl(\frac\varepsilon4-|b-r_S|\Bigr)_+\\
 &=\sum_{S\subseteq[n]:\ \sum_{i\in S}a_i=b}
 P_S\Bigl(\frac\varepsilon4-(b-r_S)\Bigr),
\end{align*}
where step (I) is by \eqref{eq:value}, in which $r_{[n]}$ has the coefficient
$1-2+1=0$, and step (II) is by \eqref{eq:second-diff} with $r=r_S$.

Let $S$ be a solution. It has $k$ elements, so
\begin{align*}
 (n+1)\binom nk P_S&=\prod_{i\in[n]\setminus S}(1-a_i)
 \in\Bigl[1-\sum_{i\in[n]\setminus S}a_i,\,1\Bigr]\subseteq[1-A,\,1],\\
 \frac4\varepsilon\Bigl(\frac\varepsilon4-(b-r_S)\Bigr)
 &\in\Bigl[1-\frac{2A^2}\varepsilon,\,1\Bigr].
\end{align*}
Both lower bounds are positive, as $A<1/2$ and by the choice of
$\varepsilon$. As
$\varepsilon\leq A$, we have $A\leq A^2/\varepsilon$, so
\begin{align*}
 1-\frac{3A^2}\varepsilon
 &\leq1-A-\frac{2A^2}\varepsilon
 \leq(1-A)\Bigl(1-\frac{2A^2}\varepsilon\Bigr)\\
 &\leq\frac{4(n+1)}\varepsilon\binom nk P_S\Bigl(\frac\varepsilon4-(b-r_S)\Bigr)
 \leq1.
\end{align*}
We sum over the $N$ solutions.
\end{proof}

\subsection{Solutions of equal size}\label{app:equal}

By the following lemma and the $\SharpP$-completeness of
\CountSS~\cite{FH09}, \CountSS{} remains $\SharpP$-complete on the
instances in which all solutions have the same number $k$ of elements.

\begin{lemma}\label{lem:equal}
A polynomial-time algorithm maps every instance of \CountSS{} with $n$ numbers
to an instance with $2n$ numbers and the same number of solutions, all of
which have $n$ elements.
\end{lemma}

\begin{proof}
Let the instance have the numbers $a_1,\dots,a_n$, with total $A$, and the
target $b$. For $i\in[n]$, the new instance has the numbers
$a'_i=2A\cdot3^{i-1}+a_i$ and $a'_{n+i}=2A\cdot3^{i-1}$, and its target is
$b'=(3^n-1)A+b$. Its numbers are positive integers with $O(n+\log A)$ bits,
and its total is $(2\cdot3^n-1)A\geq b'$.

For $U\subseteq[2n]$ and $i\in[n]$, let $t_i=|U\cap\{i,n+i\}|\in\{0,1,2\}$.
Then
\begin{align*}
 \sum_{\ell\in U}a'_\ell&=2A\sum_{i=1}^nt_i3^{i-1}+\sum_{i\in U\cap[n]}a_i,\\
 b'&=2A\sum_{i=1}^n3^{i-1}+b.
\end{align*}
If $U$ is a solution of the new instance, then
$\sum_{i\in U\cap[n]}a_i-b$ is a multiple of $2A$ in $(-2A,2A)$, and hence
zero. So $U$ is a solution of the new instance if and only if
$\sum_{i\in U\cap[n]}a_i=b$ and $\sum_it_i3^{i-1}=\sum_i3^{i-1}$. As base-$3$ representations with digits in
$\{0,1,2\}$ are unique, the second condition holds if and only if $t_i=1$ for
every $i\in[n]$. So the map $S\mapsto S\cup\{n+i:i\in[n]\setminus S\}$ is a
bijection from the solutions of the original instance to the solutions of the
new instance, and the latter have $n$ elements.
\end{proof}

\subsection{Proof of Lemma~\ref{lem:opt}}\label{app:opt}

\optlemma*

\begin{proof}
Let $\mathcal B$ be an algorithm that computes the online optimum of every
prophet secretary instance in which every value is $0$ or $1$, except for one
deterministic value in $(0,1)$. We use $\mathcal B$ as an oracle. As
\CountSS{} is $\SharpP$-complete, it suffices to compute the number $N$ of
solutions of an instance of \CountSS{} with $m$ numbers, in polynomial time
with queries to $\mathcal B$. By \cref{lem:equal}, $N$ is also the number of
solutions of an instance with $n=2m$ numbers $a_1,\dots,a_n$, total $A$ and
target $b$, whose solutions all have $k=m$ elements.

We multiply these numbers and the target by $\varepsilon=1/(2^{n+3}A^2)$.
Afterwards they are multiples of $\varepsilon$, and their total
$A/(2^{n+3}A^2)$, which we again denote by $A$, satisfies
$\varepsilon=2^{n+3}A^2>2A^2$. We query $\mathcal B$ on $I(x)$ for
$x\in\{b-\varepsilon/4,b,b+\varepsilon/4\}$, which lie in $(0,1)$ by the proof
of \cref{lem:counting}, and obtain $\Delta_1$ and $\Delta_2$. As $N\leq2^n$,
\cref{lem:counting} gives
\[
 \Bigl|\frac{4(n+1)}\varepsilon\binom nk(\Delta_1-\Delta_2)-N\Bigr|
 \leq\frac{3A^2N}\varepsilon\leq\frac{3\cdot2^nA^2}{2^{n+3}A^2}
 =\frac38<\frac12.
\]
So $N$ is the integer nearest to
$\frac{4(n+1)}\varepsilon\binom nk(\Delta_1-\Delta_2)$. All numbers in this
computation have polynomially many bits, so it runs in polynomial time.
\end{proof}

\subsection{The decision on the first arrival}\label{app:first}

\begin{lemma}\label{lem:first}
Let $I$ be a prophet secretary instance, and let $I'$ be $I$ with an added
online vertex $j_0$ of deterministic value $q\neq\OPT(I)$. If $j_0$ arrives
first, every optimal policy for $I'$ accepts $q$ if $q>\OPT(I)$, and rejects
$q$ if $q<\OPT(I)$.
\end{lemma}

\begin{proof}
Let $F$ be the event that $j_0$ arrives first. Given $F$, the online vertices
of $I$ arrive after $j_0$ in a uniformly random order. So if a policy for $I'$
rejects $q$ on $F$, its later decisions form a policy for $I$.

Let $\pi$ be a deterministic policy for $I'$ that, on $F$, rejects $q$
although $q>\OPT(I)$, or accepts $q$ although $q<\OPT(I)$. Then
$\E[M_\pi\mid F]\leq\min\{q,\OPT(I)\}$. We fix
$\delta\in(0,|q-\OPT(I)|)$, and we let $\pi'$ follow $\pi$ outside $F$. On
$F$, we let $\pi'$ accept $q$ if $q>\OPT(I)$, and otherwise reject $q$ and
follow a policy for $I$ with expected reward at least $\OPT(I)-\delta$. Then
$\E[M_{\pi'}\mid F]\geq\max\{q,\OPT(I)\}-\delta$, and
\begin{align*}
 \OPT(I')-\E[M_\pi]
 &\geq\E[M_{\pi'}]-\E[M_\pi]\\
 &=\Prb(F)\bigl(\E[M_{\pi'}\mid F]-\E[M_\pi\mid F]\bigr)\\
 &\geq\Prb(F)\bigl(\max\{q,\OPT(I)\}-\delta-\min\{q,\OPT(I)\}\bigr)\\
 &=\Prb(F)\bigl(|q-\OPT(I)|-\delta\bigr)>0.
\end{align*}
A randomized policy is a mixture of deterministic policies, each with expected
reward at most $\OPT(I')$. If the randomized policy is optimal, then with
probability one, the mixture draws an optimal deterministic policy. So on $F$,
the randomized policy accepts $q$ if $q>\OPT(I)$, and rejects $q$ if
$q<\OPT(I)$.
\end{proof}

\subsection{Proof of Theorem~\ref{thm:decisions}}\label{app:decisions}

\decisionsthm*

\begin{proof}
Let $\mathcal A$ be an algorithm that computes an optimal policy on every
prophet secretary instance in which every value is $0$ or $1$, except for two
deterministic values in $(0,1)$. We use $\mathcal A$ as an oracle. By
\cref{lem:opt}, it suffices to compute $\OPT(I)$ in polynomial time with
queries to $\mathcal A$, for every prophet secretary instance $I$ in which
every value is $0$ or $1$, except for one deterministic value $x\in(0,1)$.

Let $J$ be the set of online vertices of $I$, let $c$ be the denominator of
$x$, and for $j\in J$, let $d_j$ be the product of the denominators of the
probabilities of $X_j$, and let $D=|J|!\cdot c\cdot\prod_{j\in J}d_j$. An
arrival order and values $y_j$ of the online vertices $j\in J$ occur together
with probability
\[
 \frac1{|J|!}\prod_{j\in J}\Prb(X_j=y_j)\in\frac1{|J|!\prod_{j\in J}d_j}\,\mathbb Z.
\]
For a deterministic policy $\pi$ for $I$, the arrival order and the values
determine $M_\pi\in\{0,1,x\}\subseteq\frac1c\mathbb Z$, so $D\,\E[M_\pi]$ is an integer.
A randomized policy is a mixture of deterministic policies, and there are
finitely many deterministic policies. So $\OPT(I)=\E[M_\pi]$ for a
deterministic policy $\pi$, and $D\,\OPT(I)$ is an integer in $[0,D]$. Moreover,
$\log D$ is polynomial in the size of $I$.

For an integer $z\in[0,D)$, we add to $I$ an online vertex of deterministic
value $q=(z+1/2)/D\in(0,1)$, and we query $\mathcal A$ on this instance and
the history in which the added vertex arrives first. As $Dq$ is not an
integer, $q\neq\OPT(I)$. By \cref{lem:first}, $\mathcal A$ rejects $q$ if and
only if $\OPT(I)>q$, that is, $D\,\OPT(I)\geq z+1$. So a binary search over $z$
computes $D\,\OPT(I)$ with $O(\log D)$ queries.
\end{proof}

\end{document}